\documentclass[sigconf,nonacm]{acmart}
\usepackage{epsfig}
\usepackage[english]{babel}
\usepackage[T1]{fontenc}

\usepackage{textcomp}
\usepackage{xcolor}
\usepackage{comment}

\usepackage{amsmath}
\usepackage{amsthm}
\usepackage{amsfonts}
\usepackage{mathrsfs}
\usepackage{graphicx}
\usepackage{booktabs}
\usepackage[labelfont=bf]{caption}
\usepackage{subcaption}
\usepackage{verbatim} 
\usepackage{float} 
\usepackage[algoruled,noline,longend,linesnumbered]{algorithm2e}
\usepackage{tabularx,booktabs}
\newcolumntype{V}{>{\centering\arraybackslash}}
\newcolumntype{W}{>{\centering\arraybackslash}p{1.25cm}}
\newcolumntype{X}{>{\centering\arraybackslash}p{0.95cm}}
\newcolumntype{Y}{>{\centering\arraybackslash}p{1.15cm}X}
\newcolumntype{Z}{>{\centering\arraybackslash}p{1.05cm}X}
\DeclareMathOperator*{\E}{\mathbb{E}}

\DeclareMathOperator*{\Sspace}{\mathcal{S}}
\DeclareMathOperator*{\A}{\mathcal{A}}
\DeclareMathOperator*{\Nagents}{\mathcal{N}}

\usepackage{tikz}
\usetikzlibrary{calc} % Needed for steps figure 2
\DeclareRobustCommand{\circled}[1]{%
  {\tikz[baseline=(char.base)]{\node[shape=circle,draw,inner sep=0.6pt] (char) {\fontsize{10pt}{10pt}\selectfont #1};}}
}

\AtBeginDocument{%
  \providecommand\BibTeX{{%
    \normalfont B\kern-0.5em{\scshape i\kern-0.25em b}\kern-0.8em\TeX}}}

\begin{document}
\settopmatter{printfolios=true}
\title{GraphCC: A Practical Graph Learning-based Approach to Congestion Control in Datacenters}

\author{Guillermo Bernárdez, José Suárez-Varela, Xiang Shi, Shihan Xiao, Xiangle Cheng, Pere~Barlet-Ros and Albert~Cabellos-Aparicio}
\thanks{G. Bernárdez, P. Barlet-Ros and A. Cabellos-Aparicio are with Barcelona Neural Networking Center, Universitat Politècnica de Catalunya, Barcelona, Spain. Contact: guillermo.bernardez@upc.edu}
\thanks{J. Suárez-Varela is with Telefonica Research, Madrid, Spain.}
\thanks{X.~Shi, S. Xiao and X. Cheng are with the Network Technology Lab., Huawei Technologies Co., Ltd., Beijing, China.}
%\author{Guillermo Bernardez}
%\email{webmaster@marysville-ohio.com}
%\affiliation{%
%  \institution{Institute for Clarity in Documentation}
%  \streetaddress{P.O. Box 1212}
%  \country{USA}

%\author{Lars Th{\o}rv{\"a}ld}
%\affiliation{%
%  \institution{The Th{\o}rv{\"a}ld Group}
%  \streetaddress{1 Th{\o}rv{\"a}ld Circle}
%\email{larst@affiliation.org}


\begin{abstract}
  Congestion Control (CC) plays a fundamental role in optimizing traffic in Data Center Networks (DCN). Currently, DCNs mainly implement two main CC protocols: DCTCP and DCQCN. Both protocols ---~and their main variants~--- are based on Explicit Congestion Notification (ECN), where intermediate switches mark packets when they detect congestion. The ECN configuration is thus a crucial aspect on the performance of CC protocols. Nowadays, network experts set static ECN parameters carefully selected to optimize the average network performance. However, today's high-speed DCNs experience quick and abrupt changes that severely change the network state (e.g., dynamic traffic workloads, incast events, failures). This leads to under-utilization and sub-optimal performance. This paper presents GraphCC, a novel Machine Learning-based framework for in-network CC optimization. Our  distributed solution relies on a novel combination of Multi-agent Reinforcement Learning (MARL) and Graph Neural Networks (GNN), and it is compatible with widely deployed ECN-based CC protocols. GraphCC deploys distributed agents on switches that communicate with their neighbors to cooperate and optimize the global ECN configuration. In our evaluation, we test the performance of GraphCC under a wide variety of scenarios, focusing on the capability of this solution to adapt to new scenarios unseen during training (e.g., new traffic workloads, failures, upgrades). We compare GraphCC with a state-of-the-art MARL-based solution for ECN tuning ---~(\emph{ACC})~---, and observe that our proposed solution outperforms the state-of-the-art baseline in all of the evaluation scenarios, showing improvements up to 20\% in Flow Completion Time as well as significant reductions in buffer occupancy (38.0-85.7\%).
\end{abstract}

\keywords{Graph Neural Networks, Multi-Agent Reinforcement Learning, Congestion Control, Datacenters}

\maketitle

\section{Introduction}

The last decade has witnessed an ever-growing interest on optimizing Data Center Networks (DCN) given the critical services they run and the high CAPEX and OPEX these infrastructures entail. Among the large spectrum of traffic optimization mechanisms~\cite{li2021survey}, Congestion Control (CC) has been especially explored in the literature~\cite{alizadeh2010data,zhu2015congestion,mittal2015timely,montazeri2018homa,li2019hpcc,kumar2020swift,yan2021acc}. 
Nowadays, most production DCNs run two main CC protocols: $(i)$ DCTCP~\cite{alizadeh2010data}, for datacenters based on the traditional TCP/IP stack, and $(ii)$ DCQCN~\cite{zhu2015congestion}, for emerging RDMA-based DCNs. Both CC schemes rely on congestion notifications raised by intermediate switches in the network ---~known as the Explicit Congestion Notification (ECN) mechanism~\cite{ramakrishnan2001rfc3168}. This is the only feedback that end-hosts receive to dynamically adapt the flow rate. The ECN configuration thus become a crucial aspect for optimizing traffic in today's datacenters~\cite{shan2018ecn,yan2021acc}, and 
finding the optimal ECN parameters is a complex and time consuming task.
Nowadays, network experts end up selecting configurations that can work well on average, while being conservative enough to absorb traffic microbursts and avoid queue buildup~\cite{shan2018ecn}.

However, traffic in modern high-speed DCNs is more and more dynamic. For example, in emerging applications ---such as distributed cloud storage--- it is very frequent to find incast events (synchronous many-to-one connections), which put great pressure over specific switch ports for short time spans. At the same time, production DCNs experience failures frequently~\cite{zhou2014wcmp, liu2013f10}. This means topology changes that break a main design principle of today's DCNs: network symmetry. Failures cause network imbalance, which may lead to severe performance degradation (up to 40\% throughput reduction in real-world networks~\cite{gill2011understanding}). 

In this vein, Machine Learning (ML) has raised a special interest as a suitable technique to dynamically optimize CC in DCNs. Nowadays, we can attest some pioneering ML-based CC proposals, such as AuTO~\cite{chen2018auto}, Aurora~\cite{jay2019deep}, or Orca~\cite{abbasloo2020classic}. However, these solutions are not compatible with widely deployed equipment in datacenters, as they propose to re-implement the network stack. A more recent solution, ACC~\cite{yan2021acc}, proposes to perform in-network optimization by dynamically adapting the ECN configuration on switches. This solution has shown outstanding performance in production environments and it is compatible with current datacenter equipment running widely deployed ECN-based CC protocols (e.g., DCTCP, DCQCN). Nevertheless, ACC is designed for online training; i.e. it gradually learns how to adapt to the current network conditions. As a result, it may suffer from critical transient performance degradation when traffic changes. In general, online training is not always appropriate in production environments, as: \textit{(i)} it carries an implicit uncertainty on what would be the resulting performance of agents after re-training, \textit{(ii)} the training adds an extra execution cost, and \textit{(iii)} it may be not compatible with legacy hardware ---or simply training takes too much time due to the computational requirements for training models there.

This paper presents GraphCC, a ML-based framework for in-network CC optimization that does not need further training once deployed; by design, our solution is able to adapt to varying DCN conditions despite it being trained in a fully offline manner. Similar to ACC~\cite{yan2021acc}, GraphCC dynamically optimizes the ECN configuration on switches, and it is compatible with widely deployed CC protocols (e.g., DCTCP, DCQCN). However, our method is based on a novel combination of Multi-Agent Reinforcement Learning (MARL) and Graph Neural Networks (GNN) that, after training, produces a single agent implementation that can be deployed in a distributed way on switches to optimize the ECN configuration at the interface level. In contrast to previous proposals, deployed GraphCC agents do communicate with adjacent agents to get local context, and they actually learn how to cooperate to optimize the global Flow Completion Time~(FCT).

We evaluate GraphCC under a diverse spectrum of scenarios with DCQCN, including different real-world traffic workloads unseen during the training phase. Also, we test how this solution behaves under aggressive incast events and drastic topology changes, such as link failures or network upgrades. We compare the performance with respect to: $(i)$ a static ECN setting used in Alibaba's production networks~\cite{li2019hpcc}, and $(ii)$ ACC~\cite{yan2021acc}, the previously mentioned state-of-the-art ML-based solution for dynamic ECN tuning. In our evaluation, we use the average FCT slowdown as a reference~\cite{alizadeh2013pfabric, bai2015information, gao2015phost}.
Our experimental results show that GraphCC achieves improvements of up to 20\% in the avg. FCT slowdown w.r.t. ACC without re-training. Likewise, by tuning the ECN configuration GraphCC learns to optimize flow-level performance while keeping short queue lengths (reduction of 38.0-85.7\% w.r.t. ACC). This may be beneficial to achieve stability under unpredictable traffic microbursts~\cite{shan2018ecn}, and it is a trend already seen in other near-optimal state-of-the-art CC mechanisms relying on advanced telemetry, such as HPCC~\cite{li2019hpcc}.

Congestion Control (CC) has been extensively studied in the past. As a result, there exists a plethora of pioneering solutions for DCNs tackling the problem from different angles, such as RTT-based~\cite{mittal2015timely, kumar2020swift}, credit-based~\cite{cho2017credit,montazeri2018homa}, or telemetry-based~\cite{li2019hpcc, ben2020pint} mechanisms. Nowadays most production DCNs implement two main well-established CC protocols: DCTCP~\cite{alizadeh2010data}, and \mbox{DCQCN~\cite{zhu2015congestion}}. The former is the main standard in traditional networks based on the TCP/IP stack, while the latter is the \emph{de facto} standard in modern RDMA-based networks\footnote{RDMA stands for Remote Direct Memory Access. It is a link-level technology that optimizes memory access across distributed nodes in DCNs.}. Both state-of-the-art mechanisms \mbox{---~as} well as their enhanced schemes~\cite{zhang2019enabling,mittal2018revisiting}~--- rely on Explicit Congestion Notification (ECN)~\cite{ramakrishnan2001rfc3168}, so that switches mark packets when they experience congestion, and end-hosts dynamically adapt their transmission rate accordingly.

This paper focuses on in-network optimization of widely deployed ECN-based CC protocols (e.g., DCTCP, DCQCN). GraphCC attempts to optimize the handling of congestion notifications in switches, which is a crucial component of CC protocols to efficiently optimize traffic~\cite{yan2021acc}. In this context, both DCTCP and DCQCN implement a similar approach: switches mark the Congestion Experienced~(CE) bit of packets in case the queue length exceeds some predefined thresholds. DCTCP adopts a hard cutoff, i.e., all packets are marked when the queue length exceeds a certain value $k$. Instead, DCQCN implements a softer RED-like probabilistic approach based on three ECN configuration parameters $\{k_{min}, k_{max}, p_{max}\}$~\cite{floyd1993random}. These parameters have a significant impact on the resulting network performance (e.g., FCTs), and their optimal values are highly dependent on the current traffic conditions~\cite{yan2021acc}. This poses a great challenge on how to dynamically adapt these values to better exploit network resources, as further discussed in Section~\ref{sec:motivation}.

\begin{figure}[!t]
\centering
    \includegraphics[width=\columnwidth]{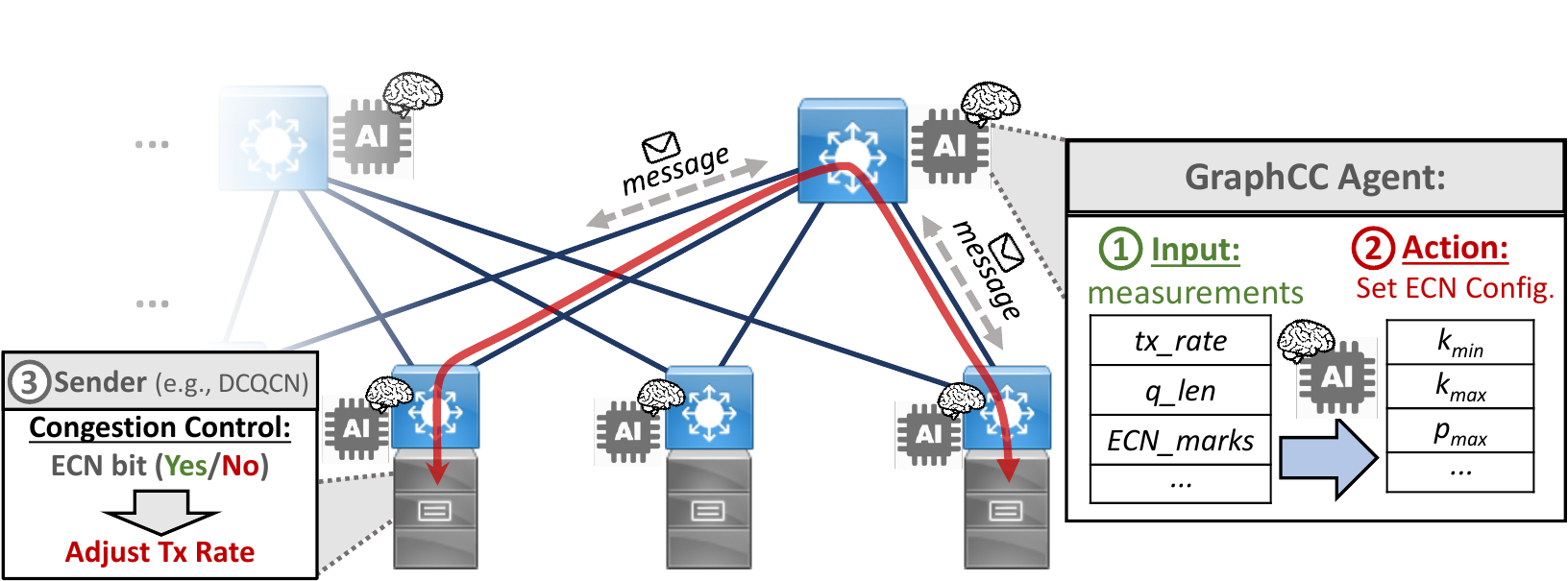}
  \caption{Schematic representation of GraphCC: \mbox{Distributed} in-network CC optimization.}
  \label{fig:scenario}
  \vspace{-0.3cm}
\end{figure} 

In this context, GraphCC deploys a set of distributed agents in switches that communicate between them to jointly optimize the ECN configuration on NICs. The operational workflow ---Figure~\ref{fig:scenario} shows a schematic representation--- is as follows:

\textbf{1) Measurement collection on NICs (Fig.~\ref{fig:scenario}; \mbox{step~\circled{1}\hspace{-2pt})}:} First, the distributed agents of GraphCC retrieve basic measurements from their local NICs. Specifically, they collect the bytes transmitted by the NIC ($tx\_rate$), the queue length ($q\_len$), and the number of packets marked by ECN in the past ($ECN\_marks$). These measurements are commonly supported by commercial switches~\cite{broadcom-products,broadcom-news, intel-tofino}, and can be locally obtained with low computational overhead at microsecond timescales~\cite{yan2021acc}.

\textbf{2) GraphCC agents optimize the ECN configuration (Fig.~\ref{fig:scenario}; step~\circled{2}\hspace{-3pt}):} Once agents collect NIC measurements, they start a communication with other agents deployed in adjacent switches to gain a local context. This communication is done through a novel NN-driven message passing, where agents exchange messages directly encoded by NN modules in order to find the best ECN settings in their associated NICs. This communication only requires to exchange few bytes with neighboring switches and can take few $\mu$s (the base link propagation delay in production DCNs is typically 1-2$\mu$s~\cite{li2019hpcc,yan2021acc}). To set the new ECN parameters ---~e.g., $\{k_{min}, k_{max}, p_{max}\}$ in DCQCN~--- agents can directly interface with forwarding chips using their API, which is typically vendor-specific. At this point, switches start to mark packets according to the new ECN settings. More details about the architecture, inner workings, communication overhead and performance of GraphCC are described in Sections~\ref{sec:solution} and~\ref{sec:evaluation}.

\textbf{3) End-hosts adjust the flow rate (Fig.~\ref{fig:scenario}; step~\circled{3}\hspace{-2pt}):} Lastly, the CC protocol executed at end-hosts (e.g., DCTCP, DCQCN) adjusts the flow transmission rate based on the ECN feedback. The process is as follows: if a host receives an ECN-marked packet, it notifies it to the sender in the corresponding ACK. When the sender receives the ACK, it re-computes the flow rate according to the protocol-specific algorithm (e.g., Additive-Increase/Multiplicative-Decrease). This CC mechanism thus enables to gradually react at one-RTT timescales ($\approx$10$\mu$s in high-speed DCNs~\cite{kumar2020swift,li2019hpcc,yan2021acc}). Note that the \mbox{in-network} optimization mechanism of GraphCC is orthogonal and complementary to the selection of the flow rate control algorithm (e.g., DCTCP, DCQCN). GraphCC is compatible with any ECN-based CC protocol and can be deployed along with any other well-established traffic optimization techniques, such as flow scheduling~\cite{alizadeh2013pfabric,bai2015information}.

\section{Main Challenges in CC Optimization}\label{sec:motivation}

This section discusses some key aspects to consider when optimizing CC protocols in high-speed DCNs, which have driven the design of GraphCC.

\subsection{Performance tradeoffs in Congestion Control} \label{subsec:cc-problem}

Nowadays, storage and computation speeds are some orders of magnitude faster than networking operations~\cite{kumar2020swift}. Hence, the network becomes the main performance bottleneck in today's datacenters. The main operational goal in DCNs is to maximize the throughput, while keeping low latency at the flow level. At the same time, current DCNs carry heavy-tailed traffic distributions, where a large amount of flows are short and time-sensitive, and a small portion of flows are long and throughput-sensitive~\cite{roy2015inside,zhu2015congestion, alizadeh2010data}. We discuss below the main tradeoffs to consider when optimizing CC in DCNs.

\subsubsection{Thoughput vs. latency}
A main challenge when operating DCNs is to keep a good compromise between the throughput for long flows and the latency for short flows. The main logic behind ECN-based schemes is to estimate network congestion based on the queue occupation on NICs; small queue length thresholds (e.g., [$k_{min}$, $k_{max}$] in DCQCN) lead to aggressively dropping packets and keeping short latency on queues, while large thresholds lead to higher bandwidth utilization at the expense of increased latency. In this context, the Flow Completion Time (FCT) metric has been widely accepted as the main performance indicator in datacenters nowadays~\cite{dukkipati2006flow}, as it unifies throughput and latency in a single metric. In particular, the \emph{FCT slowdown} metric ---which computes the ratio between the actual FCTs and the baseline FCTs if flows were sent at line-rate--- further introduces fairness across flows, and it is used in most state-of-the-art works to quantify the performance impact on applications~\cite{alizadeh2013pfabric, bai2015information, gao2015phost}
Finally, we note that applications in modern DCNs are mainly based on partition/aggregate design patterns~\cite{alizadeh2010data, li2019hpcc} where jobs are broken into small tasks and farmed out across servers, so then partial results are aggregated to produce the final output. This poses a special interest in minimizing the tail latency experienced by flows (e.g., 95/99-pct of FCT slowdown), as it often dominates the overall application performance~\cite{alizadeh2010data}. Based on this, in the design and evaluation of GraphCC we consider the \mbox{95-pct} and \mbox{99-pct} of the FCT slowdown as central performance metrics.

\subsubsection{Throughput vs stability}
Nowadays, DCNs are exposed to highly dynamic traffic patterns, such as incast events, where a large number of servers send traffic to a specific host. These patterns are very frequent in modern applications based on the partition/aggregate principle, as in every aggregation phase distributed workers send partial results in a synchronized way to aggregator nodes. Thus, beyond the canonical optimization goals of DCNs (i.e., throughput and latency), CC mechanisms need to account for stability. That is, to be prepared for rapid traffic variations (e.g., incast events, workload changes). Hence, a new tradeoff arises: conservative ECN settings often underutilize the network (i.e., less thoughput) while avoid fast queue buildup during transient incast events; on the contrary, maximizing network utilization leaves scarce headroom in buffers to absorb traffic microbursts and may lead to severe performance degradation~\cite{shan2018ecn}. In GraphCC, we introduce this tradeoff by explicitly including the queue length and the throughput ($tx_{rate}$) in the agents' reward function.

\subsection{ECN parameters are hard to optimize} \label{subsec:dynamic-cc}

\begin{figure}[!t]
\centering
    \includegraphics[width=\columnwidth]{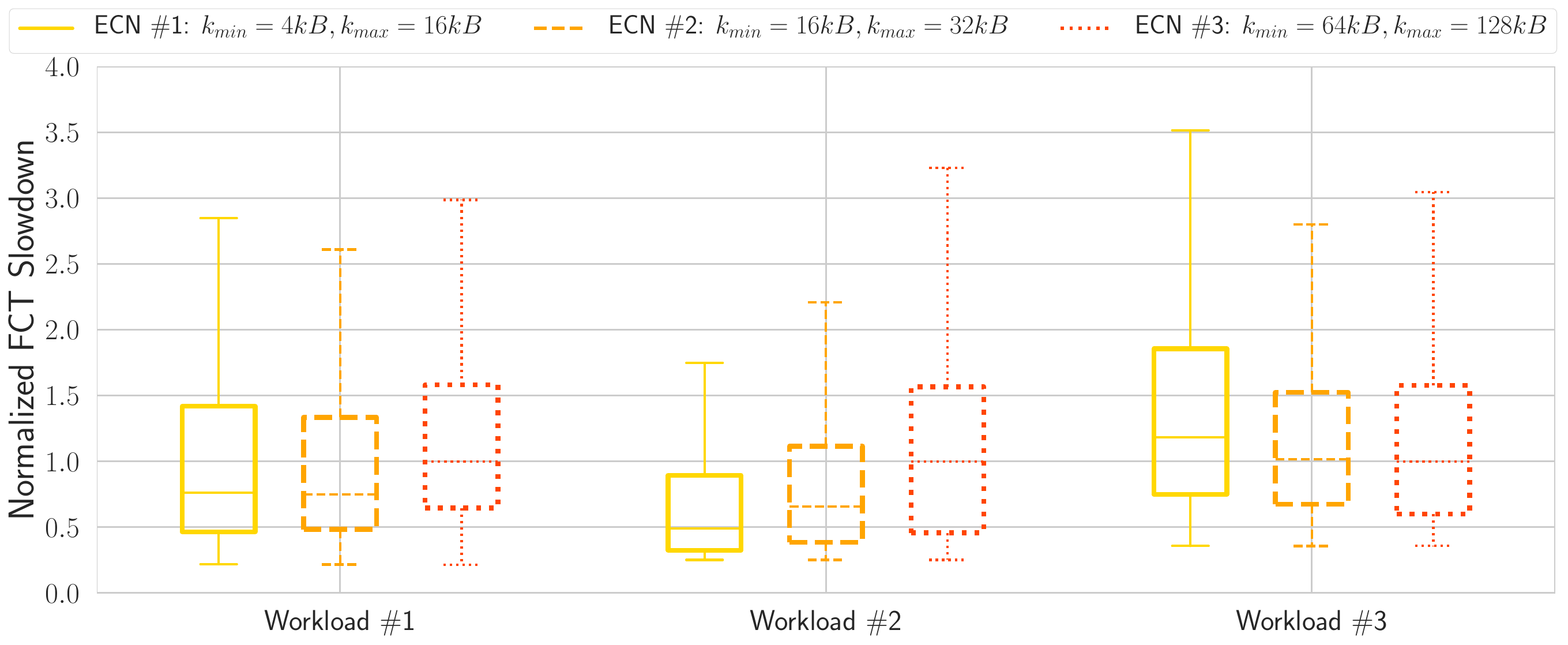}
  \caption{Normalized FCTs under various static ECN settings.}
  \label{fig:motivation}
\end{figure}

ECN settings have a large impact on network performance, and they are highly sensitive to traffic conditions~\cite{yan2021acc}. As an example, Figure~\ref{fig:motivation} shows some experimental results on a small 2-layer Clos network under three different public real-world traffic workloads, considering in each of them three different DCQCN configurations; for clarity, values are normalized by the median FCT slowdown of configuration ECN~\#3. As we can see, FCT values vary considerably depending on the workload. For example, in the case of Workload~\#1, ECN~\#1 and ECN~\#2 far outperform ECN~\#3 ($\approx$ 25\% better on median), while in Workload~\#3, ECN~\#3 outperforms the two other configurations.

As a result of the multiple optimization tradeoffs described in the previous subsection and the sensitivity of ECN settings, finding the optimal operational point in DCNs is a cumbersome task. Nowadays, network operators struggle to find good ECN parameters that can perform well on average, while at the same time guaranteeing stability under drastic events (e.g., incast, failures). This often leads to quite conservative ECN settings. The process to refine ECN parameters can typically take weeks to months~\cite{yan2021acc}, as network administrators need to make accurate workload characterizations and carefully evaluate alternative configurations under a broad casuistry (e.g., stress testing, check traffic variations, failure scenarios). More importantly, traffic workloads change drastically along the day~\cite{yan2021acc} and networks experience large daily variations on RTT (beyond 3x in real networks~\cite{zhang2019enabling}). Also, link failures occur very frequently~\cite{zhou2014wcmp, alizadeh2014conga}, which generate network asymmetries that are highly disruptive for DCNs~\cite{gill2011understanding,liu2013f10}. 
For example, the study in~\cite{gill2011understanding} states that failures may cause up to 40\% throughput reduction despite the typical link redundancy.

All this motivates the need for automatic ECN tuning techniques, such as GraphCC, that can efficiently adapt the ECN configuration to the fast traffic dynamics of nowadays DCNs. In this context ML is promising, especially Reinforcement Learning-based models ---such as GraphCC--- that can autonomously learn just by interacting with the environment. Section~\ref{sec:related-work} revisits some pioneering ML-based solutions for traffic optimization in DCNs, and further discusses their current limitations (as well as their differences w.r.t GraphCC).

\subsection{Dealing with decreasing buffer size}
There is a fundamental aspect that DCNs are increasingly witnessing as new generations of forwarding devices come to the market. During the last decade, switch capacity and link speeds have grown dramatically ($\approx$x10 in six years~\cite{goyal2022backpressure}). However, buffer size on switches have not been scaled at the same pace. This is mainly due to the high cost and technological barriers to scale switch memories to the ever-increasing link speeds. As a result, the ratio buffer size vs. switch capacity has been considerably decreased in the last years. As an example, high-end datacenter switches from Broadcom have reduced this ratio by approximately a factor of 2 in six years (from 2012 to 2018)~\cite{goyal2022backpressure}. This means that buffers now fill up faster, and this trend is expected to continue exacerbating in future datacenter generations. This poses the need for efficient CC mechanisms that can quickly react to the fast traffic dynamics (e.g., incasts). Also, keeping reduced queue length may be more important than ever to achieve stability, e.g., to have sufficient headroom to absorb traffic microbursts. In this vein, GraphCC intrinsically accounts for the minimization of the buffer occupancy (i.e., queue lengths) during the optimization process.

\section{GraphCC}\label{sec:solution}
GraphCC was carefully designed to tackle the challenges previously exposed in Section~\ref{sec:motivation}. This section describes the design and implementation of the proposed solution. We first introduce some background on its key technologies (MARL and GNN). Then, we present a mathematical formulation of GraphCC, and fully contextualize its application to the considered CC optimization scenario (see~Section~\ref{sec:scenario}). Lastly, we describe some details about the deployment of this solution.

\subsection{Background on RL and GNN}
GraphCC relies on two main ML technologies: Multi-Agent Reinforcement Learning and Graph Neural Networks. We provide an overview of key aspects of these two technologies in order to better understand the inner architecture and design choices of the proposed solution.

\subsubsection{(Multi-Agent) Reinforcement Learning} \label{subsec:back-MARL}

According to the regular setting of RL~\cite{bertsekas1996neuro}, an agent interacts with the environment in the following way: at each step $t$, the agent selects an action $a_t$ based on its current state $s_t$, to which the environment responds with a reward $r_t$ and then moves to the next state $s_{t+1}$. This interaction is modeled as an episodic, time-homogeneous Markov Decision Process (MDP) $(\Sspace, \A, r, P, \gamma)$, where $\Sspace$ and $\A$ are respectively the state and action spaces; $P$ is the transition kernel, $s_{t+1} \sim P(\cdot | s_t,a_t)$; $r_t$ represents the immediate reward given by the environment after taking action $a_t$ from state $s_t$; and $\gamma \in (0,1]$ is the discount factor used to compute the return $G_t$, defined as the --discounted-- cumulative reward from a certain time-step $t$ to the end of the episode $T$: $G_t = \sum_{t=0}^T \gamma^t r_t$. The behavior of the agent is described by a policy $\pi: \mathcal{S} \to \mathcal{A}$, which maps each state to a probability distribution over the action space, and the goal of an RL agent is to find the optimal policy in the sense that the actions it takes maximize the expected return~$\hat{G}_t$.
From among the wide variety of existing RL approaches and algortihms to tackle this problem formulation, we implemented \mbox{Q-learning}~\cite{sutton2018reinforcement}, an action-value method where the policy $\pi$ is indirectly defined from the learned estimates of the action value function $Q^\pi (s,a)~=~\E_{\pi} \left[ G_t | s_0=s, a_0=a \right]$. We also considered several extensions, such as Double DQN~\cite{hasselt2010double} and Prioritized Replay~\cite{schaul2015prioritized}, that enhance the basic \mbox{Q-learning} algorithm.

In a MARL framework there is a set of agents $\mathcal{V}$ interacting with a common environment that have to learn how to cooperate to pursue a common goal. In contrast to single-agent RL, such a setting is generally formulated as a Decentralized Partially Observable MDP (Dec-POMDP)~\cite{foerster2018deep} where, besides the global state space $\Sspace$ and action space $\A$, it distinguishes local state and action spaces for every agent --i.e., $\Sspace_v$ and $\A_v$ for $v \in \mathcal{V}$. At each time step~$t$ of an episode, each agent may choose an action $a_v^t \in \A_v$ based on local observations of the environment encoded in its current state $s_v^t \in \Sspace_v$. Then, the environment produces individual rewards $r_v^t$ (and/or a global reward $r^t$), and it evolves to a next global state $s^{t+1}\in \Sspace$ --i.e., each agent~$v$ transitions to the following state $s_v^{t+1} \in \Sspace_v$. Typically, a MARL system seeks for the optimal global policy by learning a set of local policies $\{\pi_{\theta_v}\}_{v \in \mathcal{V}}$. For doing so, most state-of-the-art MARL solutions implement traditional (single-agent) RL algorithms on each distributed agent, while incorporating some kind of cooperation mechanism between them~\cite{foerster2018deep}. 
The standard approach for obtaining a robust decentralized execution, however, is based on a centralized training where extra information can be used to guide agents' learning~\cite{oliehoek2008}.

\subsubsection{Graph Neural Networks} \label{subsec:back-GNN}

 These models are a recent family of neural networks specifically conceived to operate over graph-structured data~\cite{scarselli2008graph, battaglia2018relational}. In their basic form, they consist in associating some initial states to the different elements of an input graph, and combine them considering how these elements are connected in that graph. The resulting state representations, which now may encode some topological awareness, are then used to produce the final output of the GNN, which can be at the level of graph elements, or at a global graph level. 
 
 Among the numerous GNN variants developed to date~\cite{wu2020comprehensive}, we focus on Message Passing Neural Networks (MPNN)~\cite{gilmer2017neural}, which is a well-known type of GNN whose operation is based on an iterative message-passing algorithm that propagates information between elements in a graph \mbox{$\mathcal{G} = (\Nagents, \mathcal{E})$}. Focusing on the set of nodes, the process is as follows: 
first, each node $v \in \Nagents$ initializes its hidden state $h_v^0$ using some initial features already included in the input graph. At every message-passing step $k$, each node $v$ receives via messages the current hidden state of all the nodes in its neighborhood $\mathcal{B}(v)=\{u \in \Nagents | \exists e \in \mathcal{E}, e = (u,v) \lor e=(v,u) \}$, and processes them individually by applying a message function \textit{m(·)} together with its own internal state $h_v^k$. Then, the processed messages are combined by an aggregation function \textit{a(·)}, $M_v^k = a( \{ m(h_{v}^k, h_{i}^{k}) \}_{i \in \mathcal{B}(v)})$. Finally, an update function \textit{u(·)} is applied to each node $v$; taking as input the aggregated messages $M_{v}^{k}$ and its current hidden state $h_v^k$, it outputs a new hidden state for the next step ($k+1$), $h_{v}^{k+1} = u(h_{v}^{k}, M_{v}^{k})$.
After a certain number of message passing steps $K$, a readout function \textit{r(·)} takes as input the final node states $h_v^K$ to produce the final output of the GNN model. This readout function can predict either features of individual elements (e.g., a node's class) or global properties of the graph.
Note that a MPNN model generates \textit{a single set of message, aggregation, update, and readout functions that are replicated at each selected graph element}.

\subsection{Framework Formalization} \label{subsec:framework}

\begin{figure*}[!t]
\centering
    \includegraphics[width=1.95\columnwidth]{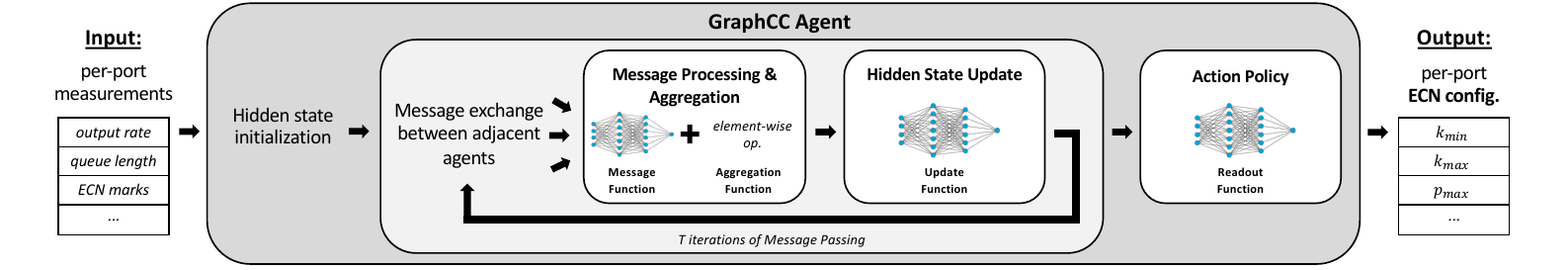}
  \caption{Operational workflow of a GraphCC agent.}
  \label{fig:flowchart}
\end{figure*} 

GraphCC models a DCN as a graph \mbox{$\mathcal{G} = (\Nagents, \mathcal{E}, \mathcal{V})$}, with $\Nagents$ and $\mathcal{E}$ being the set of nodes and edges, respectively, and $\mathcal{V}$ being the set of agents that can control some of the graph entities (nodes or edges). 
Let $\Sspace$ and $\A$ represent the global state and action spaces, respectively, defined as the joint and union of the respective agents' local spaces, $\Sspace = \prod_{v\in \mathcal{V}} \Sspace_v$ and $\A = \bigcup_{v\in \mathcal{V}} \A_v$.
By implementing Q-learning, our model aims to learn the state-action value function $Q_\theta$ for every agent $v \in \mathcal{V}$.

GraphCC makes agents learn that state-value function approximator in a fully distributed fashion --i.e., all agents end up constructing and using the very same $Q_\theta$. In fact, each agent $v \in \mathcal{V}$ computes this function mainly through message communications with their direct neighboring agents $\mathcal{B}(v)$ and their local computations, no longer needing a centralized entity responsible for collecting and processing all the global information together. Such a decentralized, message-based generation of the global function is achieved by modeling $Q_\theta$ with a MPNN (see Sec.~\ref{subsec:back-GNN}). Thus, all agents $v \in \mathcal{V}$ deployed in the network are actually \textit{replicas} of the MPNN modules (message, aggregation, update and readout functions) that perform regular message exchanges with their neighbors $\mathcal{B}(v)$ following the message passing iteration procedure of MPNNs. Note that such \textit{parameter sharing} implies that all agents share as well the same local state and action spaces. 
This reinterpretation of a MPNN as a set of copies of its internal modules is especially important due to the fact that in our approach we directly map the graph $\mathcal{G}$ to the real DCN scenario, deploying copies of MPNN modules along switches and making message communications through the network infrastructure. Hence, GraphCC naturally distributes the execution of the MPNN, and consequently is able to fully decentralize the execution pipeline. 

We formalize the optimization problem as a Dec-POMDP of standard MARL~\cite{foerster2018deep}, where all agents can act simultaneously at each step  of the optimization process. More in detail, at each time-step $t$ of an episode of length $T$, the MPNN-driven procedure of approximating $Q_\theta(s^t_v,a^t_v)$ for each agent $v \in \mathcal{V}$ \mbox{--where} $s^t_v \in \Sspace_v$ and $a^t_v\in \A_v$ refer to the local state and action at $t$-- first constructs a hidden state $h_v$ encoding meaningful information for each agent $v\in\mathcal{V}$. Each hidden state $h_v$ is updated based on the hidden representations of the neighboring agents $\mathcal{B}(v)$ ---~i.e., the node's local context~---, and its initialization $h_v^0$ is a function of the current agent state $s^t_v\in \Sspace_v$, which is in turn based on some predefined internal agent features $x_v^t$ (e.g. port utilization , queue length, etc...; more details on how agents construct their corresponding input feature vector in next section). Those hidden states are shaped during $K$ message-passing steps, where they are iteratively propagated through the graph via messages between direct neighbors. In particular, successive hidden states $h_v^k$, where $k$ accounts for the message-passing step, are computed by the message, aggregation and update functions of the MPNN, as previously described in Section \ref{subsec:back-GNN}. 

Once agents generate their final hidden representation, the readout function --following the MPNN nomenclature-- is applied separately to each agent to finally obtain its corresponding state-value function. For each agent $v\in\mathcal{V}$, the readout function takes the final hidden state $h^K_v$ as input, and produces the \mbox{q-value} \mbox{estimates} $Q_\theta(s^t_v,a_v)$ for every possible action \mbox{$a_v\in \A_v$.} Finally, following the standard procedure of \mbox{Q-learning} algorithms, at that time-step $t$ each agent selects the action with the maximum associated q-value, $a_v^t = \text{arg max}_{a_v} Q_\theta(s^t_v,a_v)$.

\subsection{Applying GraphCC to Congestion Control in DCNs}

A DCN can be described in terms of its hardware devices $\mathcal{W} = \{\mathcal{W}_s, \mathcal{W}_h \}$ ---where $\mathcal{W}_s$ and $\mathcal{W}_h$ denote the sets of switches and hosts, respectively--- and the link connections between them, $\mathcal{L} = \{ l=(w^{src},w^{dst}),  w^{src}\in\mathcal{W} \text{ and } w^{dst} \in \mathcal{W} \text{ connected}\}$. Therefore, in our graph-based model \mbox{$\mathcal{G} = (\Nagents, \mathcal{E}, \mathcal{V})$} of a DCN, we can directly relate the set of nodes $\Nagents$ with hardware devices $\mathcal{W}$, and the set of edges $\mathcal{E}$ with the actual set of network interfaces $\mathcal{L}$. GraphCC identifies each egress port of a switch as an agent, i.e. $\mathcal{V} = \{ (n^{src},n^{dst})\in \mathcal{E}, n^{src} \in \mathcal{W}_s \}$, which in particular allows to define the ECN marking thresholds at a link granularity (i.e., at the interface level)\footnote{Note that, in practice, port-based agents are deployed and run in their adjacent switches.}. In this context, we can differentiate two different neighborhoods for each link-based agent $v=(n^{src}_v,n^{dst}_v) \in \mathcal{V}$: 
 \begin{itemize}
     \item Ingress neighborhood $\mathcal{B}_i(v)$, defined as the set of links $e=(n^{src}_e,n^{dst}_e) \in \mathcal{E}$ that can potentially inject traffic into $v$, $\mathcal{B}_i (v) = \{ e\in \mathcal{E} | n^{dst}_e =n^{src}_v \}$.
     \item Egress neighborhood $\mathcal{B}_e(v)$, consisting in the links that can potentially receive traffic from $v$, $\mathcal{B}_e (v) = \{ e\in \mathcal{E} | n^{src}_e =n^{dst}_v \}$.
 \end{itemize}
 
Let us now fully describe the operation of GraphCC for CC in DCNs (Figure \ref{fig:flowchart} shows a schematic operational workflow). We assume that our solution interacts with the environment (i.e. the network) every time interval $\Delta t$, which is pre-defined. This enables to make the time evolution discrete, and hence facilitates the episodic formulation of RL. At each step $t$ of an episode, each agent $v \in \mathcal{V}$ first gathers three relevant NIC-level metrics available at the switch: ($i$) the port utilization $u_v^t$ (computed as the $tx_{rate}$ normalized by link capacity), ($ii$) the instantaneous queue length $q_v^t$, and ($iii$) the ECN marking rate $ECN_v^t$ (normalized by the link capacity). Then, agents construct their corresponding input feature vector $x_v^t$ based on the current values of these metrics as well as the values of the previous $p$ steps, i.e. 
$$x_v^t = (u_v^t,q_v^t,ECN_v^t,u_v^{t-1},q_v^{t-1},ECN_v^{t-1},\dots,u_v^{t-p},q_v^{t-p},ECN_v^{t-p}).$$ 
Empirically we found that $p=2$ works best for providing some temporal context while keeping a low input dimension.

Agents initialize their initial MPNN-based hidden state $h_v^0$ with their input feature vector $x_v^t$; since the hidden state vector dimension is typically equal or higher than that of the input feature vector, remaining components are simply 0-padded. GraphCC then executes $K$ message-passing steps in which distributed communications between agents are involved. More in detail, at each message-passing step $k$ each agent $v \in \mathcal{V}$ first sends its current hidden representation $h^k_v$ to its egress neighbors $\mathcal{B}_e(v)$, and consequently they receive a set of messages from their ingress adjacent agents $\mathcal{B}_i(v)$. At that point, agents individually combine each of the received hidden states with their own through the message function --in our case, a feed-forward neural network--, and all the resulting outputs are in turn merged into a fixed-size representation $M_v^k$ via the aggregation function. For doing so, we implement element-wise operations, such as min and max. Finally, each agent applies the update function --another fully-connected neural network-- to the aggregated information $M_v^k$ and its own hidden state $h^k_v$, which outputs its new hidden representation $h^{k+1}_v$. 

After concluding the $K$ message-passing steps, all agents end up with a final hidden state $h_v^K$ for that specific time-step $t$. This final representation is then fed by each agent into their readout module, which provides with the final values that are used to define their individual policies. In particular, since GraphCC implements a Q-learning based pipeline, each agent's readout directly outputs the q-value estimates of all possible actions for the current state of the DCN, and as described in previous subsection \ref{subsec:framework} each agent takes the action with the highest value. 
This lead us to the actual definition of the action space. GraphCC faces the CC problem by optimizing the ECN marking thresholds, and by design it is able to adapt the ECN parameters $\{ k_{min},k_{max},p_{max} \}$ for each individual switch egress port. In our implementation, we discretize the values of these parameters into some pre-defined values; we provide more details in Section \ref{subsec:setup}.

Regarding the learning process, all MPNN internal modules (message, update and readout functions, which are replicated among all agents) are trained based on the rewards computed at each step $t$ of a training episode. More in detail, in our implementation we define the reward $r_v^t$ of agent $v\in\mathcal{V}$ at step $t$ as 
$$r_v^t = w_1 \cdot f(q_v^t) + w_2 \cdot u_v^t,$$
where $f(\cdot)$ is a descreasing function with respect to the queue length $q_v^t$ of the associated port ($\text{dom}(f)=[0,1]$, based on \cite{yan2021acc}), $u_v^t$ accounts for the utilization of that port, and $w_1,w_2 \in [0,1]$ are the corresponding weights, with $w_1 + w_2 = 1$ ($w_1=0.7$ and $w_2=0.3$ worked best in our experiments). At each time step $t$ of a training episode, GraphCC gathers the global RL-based sequence $(\{x_v^t\}_{v\in\mathcal{V}},\{r_v^t\}_{v\in\mathcal{V}},\{x_v^{t+1}\}_{v\in\mathcal{V}})$ and stores it as a single sample in a replay buffer. Then our model randomly selects a batch of these samples from the buffer and performs the training and update of $Q_\theta$ accordingly~\cite{sutton2018reinforcement}.

\subsection{Discussion on Deployability}

In this section we discuss some relevant practical implications when deploying the proposed solution in DCNs. GraphCC naturally distributes the modules of the MPNN among the switches of the DCN, which enables to parallellize all the node-level computations and communications on the local neighborhood. By construction, the GNN-based modules behind GraphCC --interpreted as agents-- can be replicated and deployed in any switch, to optimize the ECN configuration on a particular NIC, regardless of the size and shape of the DCN topology considered. The process of GraphCC to exchange messages between neighbors --i.e. share the agent's hidden states-- is in fact generic and scale-invariant~\cite{ruiz2020graphon}. This provides GraphCC with excellent scalability and generalization capabilities.

GraphCC applies parameter sharing over all the NIC-based agents; thus during training the MPNN modules jointly learn from the individual perspective of all NICs in the network, and also in this process agents learn what information to exchange with neighbors in order to achieve effective coordination between agents at different levels of granularity --following the topology structure from the local context (i.e., direct neighbors) to a more global context within the network. Note, however, that no matter all agents are in fact replicas, at execution time each of them is able to specifically adjust its behaviour based on its local state and the information received from its neighbors.

Finally, the multi-agent formulation of GraphCC facilitates exploring the large solution space. By enabling the adjustment of ECN parameters $\{k_{min},k_{max},p_{max}\}$ at the interface level, the combinatorial of all the possible actions would explode from a single-agent perspective --specially taking into account that modern datacenters may have up to tens of thousands of servers~\cite{roy2015inside, singh2015jupiter}. In this sense, GraphCC approach of defining a policy for each of these NIC instances separately allows to effectively deal with such complexity and fully distribute the agent's decision making.

\section{Evaluation}\label{sec:evaluation}

\begin{figure*}[!t]
    \begin{subfigure}[]{0.32\textwidth}
	\includegraphics[width=1.0\linewidth]{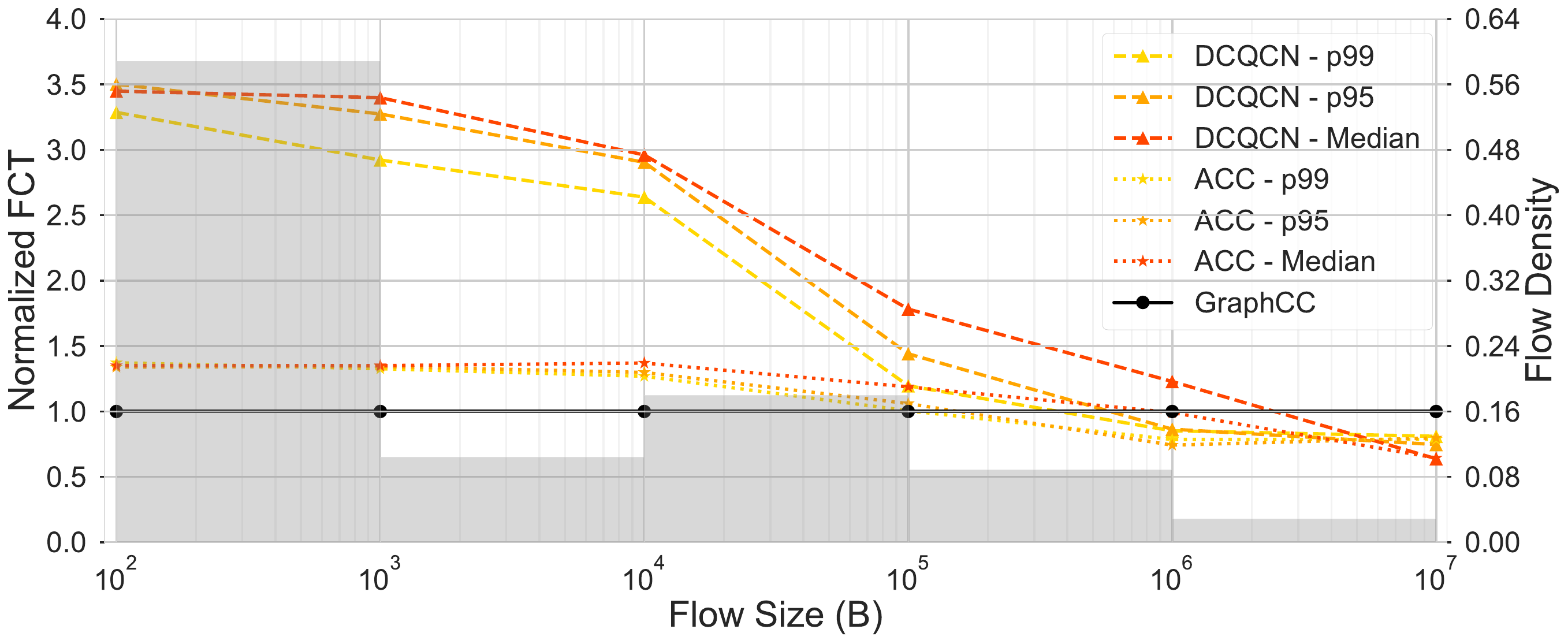}
	\vspace{-0.5cm}
        \caption{\emph{FB\_Hadoop} with 16:1 incasts}
	\label{subfig:fb_incasts}
    \end{subfigure}
    \addtocounter{subfigure}{1}
    \begin{subfigure}[]{0.32\textwidth}
	\includegraphics[width=1.0\linewidth]{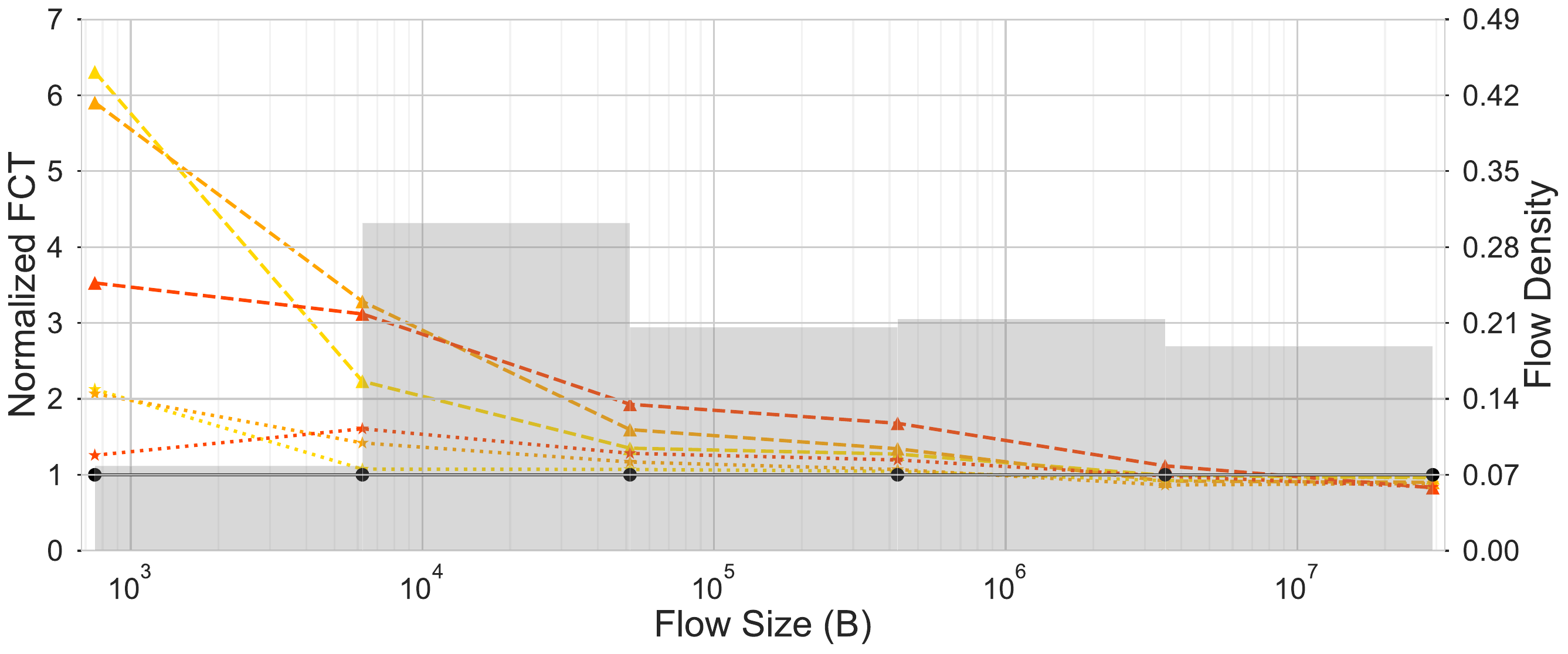}
	\vspace{-0.5cm}
        \caption{\emph{WebSearch} with 16:1 incasts}
	\label{subfig:websearch_incasts}
    \end{subfigure}
    \addtocounter{subfigure}{1}
    \begin{subfigure}[]{0.32\textwidth}
	\includegraphics[width=1.0\linewidth]{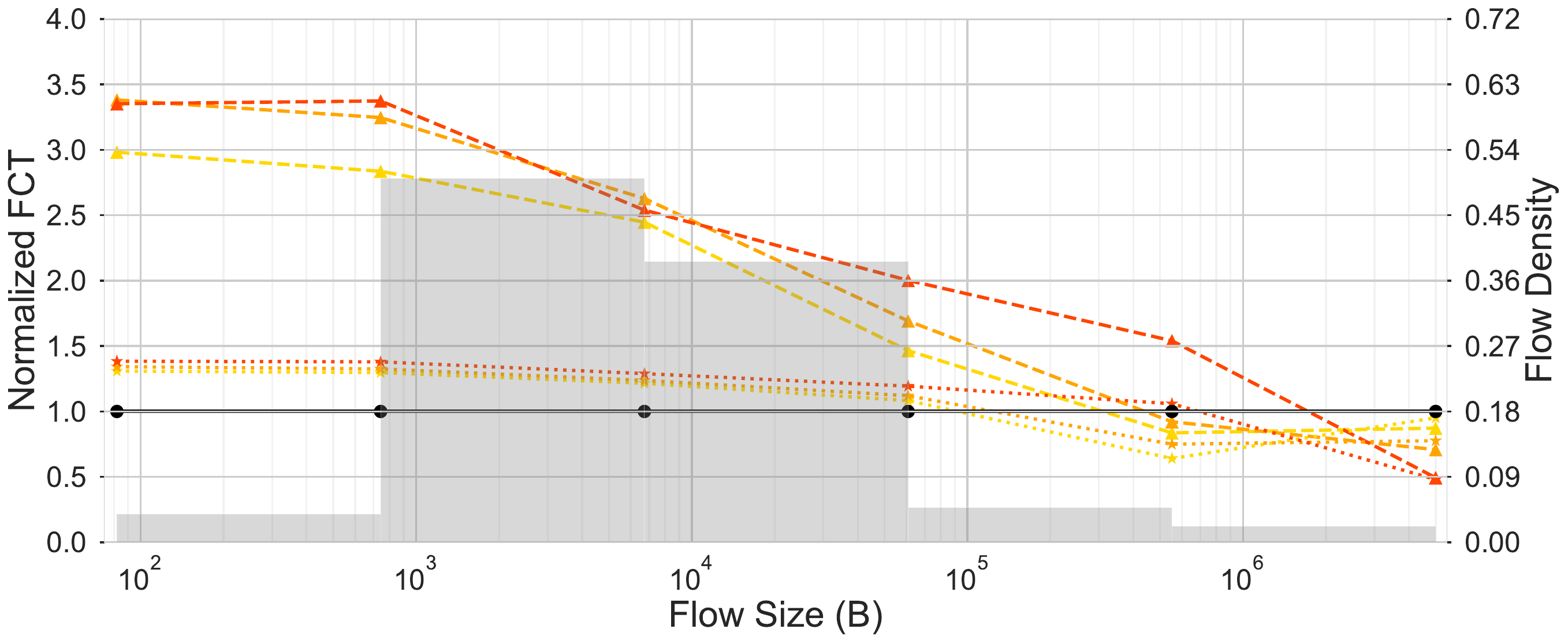}
	\vspace{-0.5cm}
        \caption{\emph{AliStorage} with 16:1 incasts}
	\label{subfig:ali_incasts}
    \end{subfigure}
    \addtocounter{subfigure}{-4}
    \begin{subfigure}[]{0.32\textwidth}
	\includegraphics[width=1.0\linewidth]{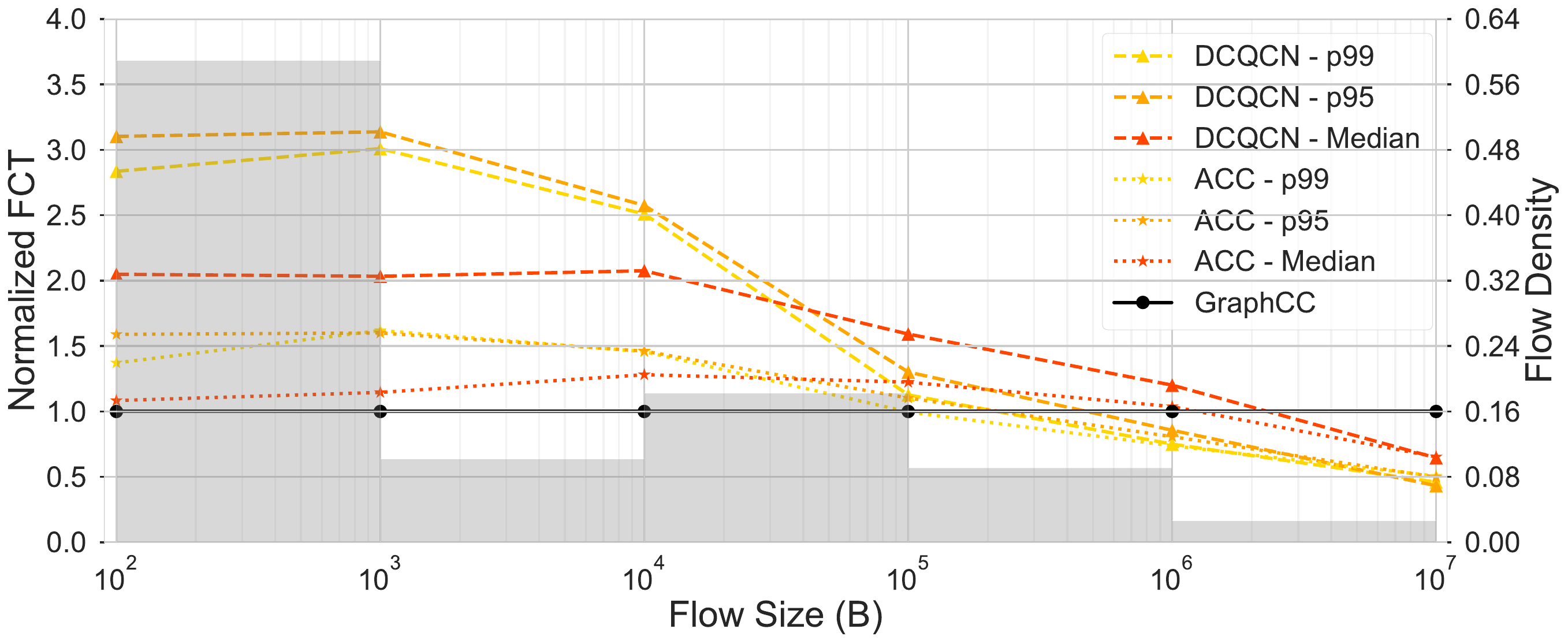}
	\vspace{-0.5cm}
        \caption{\emph{FB\_Hadoop} without incasts}
	\label{subfig:fb_no_incasts}
    \end{subfigure}
    \hspace{0.17cm}
    \addtocounter{subfigure}{1}
    \begin{subfigure}[]{0.32\textwidth}
	\includegraphics[width=1.0\linewidth]{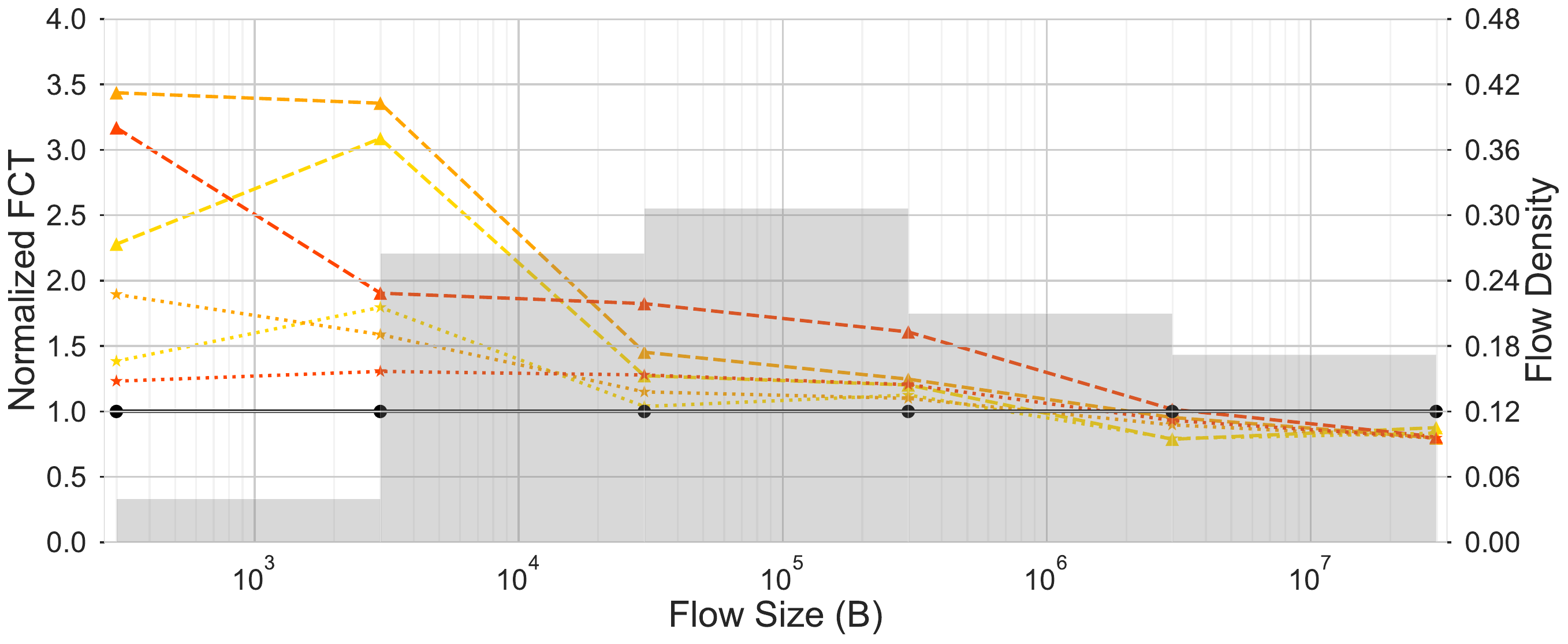}
	\vspace{-0.5cm}
        \caption{\emph{WebSearch} without incasts}
	\label{subfig:websearch_no_incasts}
    \end{subfigure}
    \hspace{0.1cm}
    \addtocounter{subfigure}{1}
    \begin{subfigure}[]{0.32\textwidth}
	\includegraphics[width=1.0\linewidth]{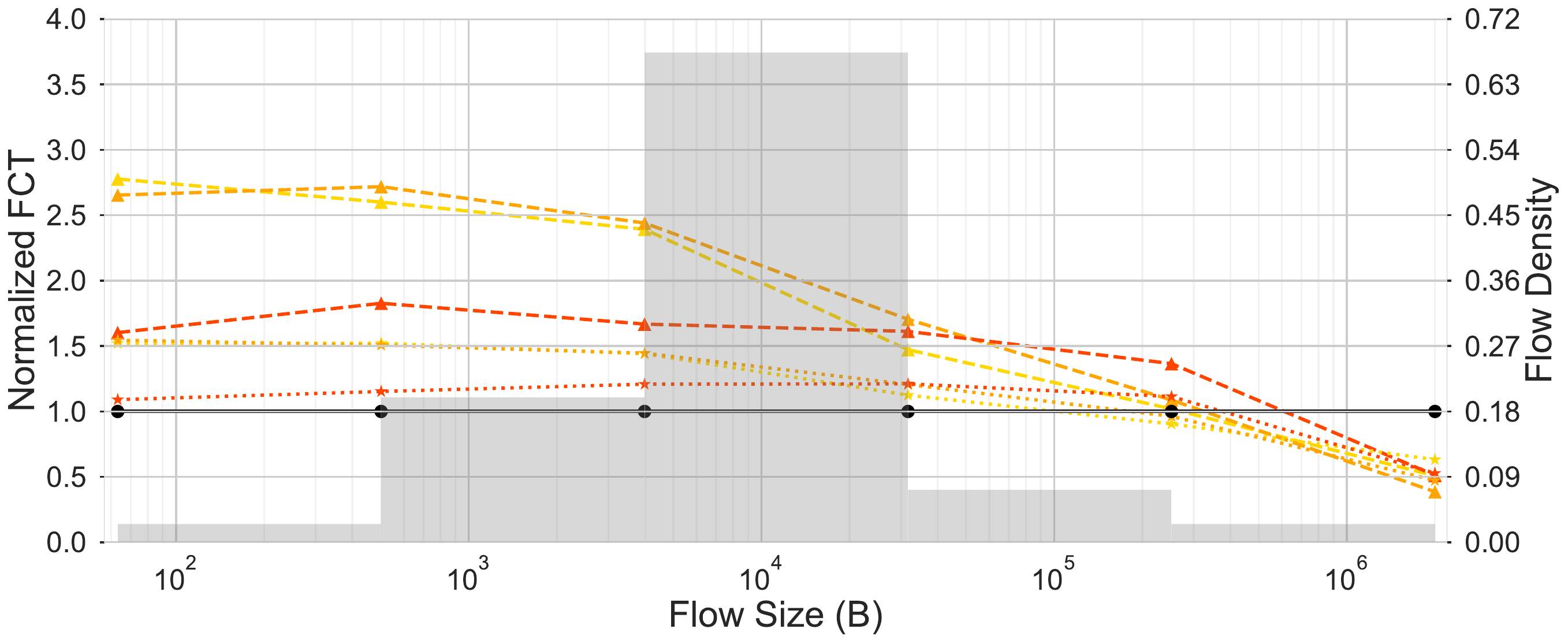}
	\vspace{-0.5cm}
        \caption{\emph{AliStorage} without incasts}
	\label{subfig:ali_no_incasts}
    \end{subfigure}
     \caption{Normalized FCT median, p95 and p99 of the DCQCN and ACC baselines --with respect to GraphCC-- when considering different traffic workloads (\emph{FB\_Hadoop}, \emph{WebSearch}, \emph{AliStorage}) with 16:1 incasts (top) or without (bottom). Our GraphCC model is exclusively trained on traffic from \emph{FB\_Hadoop} with incasts. The measured flow density can be visualized to facilitate the interpretation of the results.}
  \label{fig:websearch_and_ali}
\end{figure*}

This section comprises an extensive performance evaluation of GraphCC, including a direct comparison against ACC~\cite{yan2021acc}, a state-of-the-art ML-based solution for in-network optimization of ECN parameters. In particular, our evaluation seeks to answer the following questions:

\begin{itemize}
\item Can GraphCC accommodate different traffic patterns (e.g., incasts) and workloads unseen during training? (Sections~\ref{subsec:performance} and~\ref{subsec:workloads})
\item How does GraphCC behave under increasing network loads? E.g., up to 80\%
(Section~\ref{subsec:traffic-loads})
\item How does GraphCC perform under topology changes? E.g., failures, upgrades
(Section~\ref{subsec:topology-changes})
\end{itemize}

\subsection{Experimental setup} \label{subsec:setup}

We use the ns-3 simulator~\cite{riley2010ns}. Experiments are done in a 2-layer Clos Network similar to the testbed used in ACC~\cite{yan2021acc}, with 24 hosts, 4 leaf switches, and 2 spine switches. Switch-to-switch links have a capacity of 100Gbps, and host-to-switch links have 25Gbps. All links have a propagation delay of 1$\mu$s, consequently the maximum base RTT is 8$\mu$s. Switches have a shared-memory buffer of 32MB, which is derived from real devices~\cite{broadcom-products}. The network comprises a single RDMA domain, and PFC is enabled on switches~\cite{pfc}. End-hosts implement a DCQCN~\cite{zhu2015congestion} distribution mimicking the implementation of Mellanox ConnectX-4 cards~\cite{li2019hpcc}, with a fixed window that limits the inflight bytes to the maximum Bandwidth-Delay Product (BDP) in the network. This configuration is known to perform better than the original one, as it avoids PFC storms~\cite{li2019hpcc}. During our simulations, GraphCC agents act every 100$\mu$s, and each episode lasts 25ms. We then leave sufficient time to ensure the same set of flows finishes in all experiments, including the longest flows. 

We note that we will always consider the same GraphCC model along all our evaluation, which was exclusively trained using a real-world \emph{FB\_{Hadoop}} workload~\cite{roy2015inside} trace with a normal traffic load of 60\%~\cite{li2019hpcc} and considering periodic 16:1 incasts events~\cite{yan2021acc}. In our implementation of $Q_\theta$, message, update and  readout functions are 2-layer feed forward NNs, and the aggregation function combines an element-wise \textit{min} and \textit{max}. Regarding the tuning of some important hyper-parameters of the model\footnote{A detailed description of the rest of parameters involved will be provided with the public release of our implementation.}, the dimension of the agents' hidden states $h_v$ is set to $24$, and $K=2$ message passing steps are considered ---experimentally, we have found that more iterations do not lead to better performance, since the diameter of DCNs is typically very limited (2-3 hops). In addition, the possible ECN configurations are discretized according to the sets $k_{min}=\{2,4,8,16,32\}$, $k_{max}=\{ 16,32,64,128,256 \}$ and $p_{max} = \{0.01, 0.25, 0.5, 0.75, 1.0\}$, resulting in an agent action space $\mathcal{A}_v$ with $120$ effective combinations.

\begin{table*}[ht]
\centering
\begin{tabularx}{\textwidth}{l W *{4}{YYYZ} }
\toprule
                            &  & \multicolumn{3}{c}{\textbf{Mean FCT Slowdown}} & &  \multicolumn{3}{c}{\textbf{Mean Throughput (Mbps)}} & & \multicolumn{3}{c}{\textbf{Mean queue length (kB)}} &  \\ \cmidrule(lr){3-5} \cmidrule(l){7-9} \cmidrule(l){11-13}
                             & & \textbf{DCQCN}  &   \textbf{ACC}   & \textbf{GraphCC} &   &\textbf{DCQCN}  &   \textbf{ACC}   & \textbf{GraphCC} &   & \textbf{DCQCN}  &   \textbf{ACC}   & \textbf{GraphCC} &    \\ \midrule
Incasts  & \emph{FB\_Hadoop}         &  5.76     &  3.25   &   \textbf{2.84} & \mbox{\hspace{-0.40cm}$\rightarrow$ -12.6\%}    &  \textbf{401}     &  399     &   398 & \mbox{\hspace{-0.40cm}$\rightarrow$ -0.25\%}    &   42.4    &   12.6    &   \textbf{6.00} & \mbox{\hspace{-0.40cm}$\rightarrow$ -52.4\%}    \\
16:1 & \emph{WebSearch}         &   6.48    &   5.38   &   \textbf{5.30} & \mbox{\hspace{-0.40cm}$\rightarrow$ -1.48\%}   &   \textbf{395}    &   394    &   390 & \mbox{\hspace{-0.40cm}$\rightarrow$ -1.02\%}    &   37.3    &   9.64    &   \textbf{1.65} & \mbox{\hspace{-0.40cm}$\rightarrow$ -82,9\%}  \\
 & \emph{AliStorage}     &  5.95     &   3.49   &    \textbf{3.10} & \mbox{\hspace{-0.40cm}$\rightarrow$ -11.2\%}  &    399   &   \textbf{400}    &   399 & \mbox{\hspace{-0.40cm}$\rightarrow$ -0.25\%}    &   49.2    &   16.7    &   \textbf{9.70} & \mbox{\hspace{-0.40cm}$\rightarrow$ -41,9\%} 
\\ \midrule
No & \emph{FB\_Hadoop}         & 3.94  &  2.59  & \textbf{2.16} & \mbox{\hspace{-0.40cm}$\rightarrow$ -16.6}\% & \textbf{362} &    361   &  356 & \mbox{\hspace{-0.40cm}$\rightarrow$ -1.38\%}  &  26.6     &   10.0 &   \textbf{3.83} & \mbox{\hspace{-0.40cm}$\rightarrow$ -61.7}\%   \\
Incasts  & \emph{WebSearch}         &   4.37   &  3.70    &  \textbf{3.65} & \mbox{\hspace{-0.40cm}$\rightarrow$ -1.35\%}    &    345   &  \textbf{346}    &   343 & \mbox{\hspace{-0.40cm}$\rightarrow$ -0.86\%}   &  24.2   &   7.95    &   \textbf{1.14} & \mbox{\hspace{-0.40cm}$\rightarrow$ -85,7\%}   \\
 & \emph{AliStorage}     &  3.72   &  2.71  & \textbf{2.38} &  \mbox{\hspace{-0.40cm}$\rightarrow$ -12.2\%}  &   \textbf{355} &   \textbf{355}    &   354 &\mbox{\hspace{-0.40cm}$\rightarrow$ -0.28\%}  &   27.0    &    13.2   &   \textbf{6.73} & \mbox{\hspace{-0.40cm}$\rightarrow$ -49.0\%}  \\ \midrule
Big  & \mbox{High (70\%)}         &  5.81     &  3.41   &   \textbf{2.80} & \mbox{\hspace{-0.40cm}$\rightarrow$ -17.8\%}    &  \textbf{416}     &  415     &   411 & \mbox{\hspace{-0.40cm}$\rightarrow$ -0.96\%}    &   42.0    &   12.5    &   \textbf{5.34} & \mbox{\hspace{-0.40cm}$\rightarrow$ -57.2\%}    \\
loads & \mbox{Extreme (80\%)}         & 7.01  &  4.27  & \textbf{3.41} & \mbox{\hspace{-0.40cm}$\rightarrow$ -20.1}\% & \textbf{444} &    443   &  438 & \mbox{\hspace{-0.40cm}$\rightarrow$ -1.13\%}  &  50.8     &   15.4 &   \textbf{6.38} & \mbox{\hspace{-0.40cm}$\rightarrow$ -58.6}\%   \\
 \bottomrule
\end{tabularx}
\caption{Mean FCT slowdown, throughput, and queue length obtained by GraphCC and the baselines for different traffic workloads (\emph{FB\_Hadoop}, \emph{WebSearch}, \emph{AliStorage}) ---with and without incasts---, and with high traffic loads. It also includes the relative difference (in \%) of GraphCC results w.r.t. those of ACC.}
\label{tab:metrics_incasts}
\end{table*}

\subsection{Benchmarks and Performance Metrics}
\label{subsec:benchmarks}

Before presenting our experimental results, in this section we aim to define how we actually quantify the performance of GraphCC. First of all, we considered as baselines the following two state-of-the-art methodologies ---revisited in Section \ref{sec:related-work}--- compatible with widely-deployed ECN protocols:

\begin{itemize}
    \item  DCQCN~\cite{zhu2015congestion}: This benchmark represents the most widely used approach in today's DCNs, i.e., careful selection of static ECN parameters. It implements the static ECN configuration used in Alibaba's production DCNs~\cite{li2019hpcc}:\\ $k_{min} = 100KB$ × $\frac{Bw}{25Gbps}$; $k_{max} = 400KB$ × $\frac{Bw}{25Gbps}$. 
    \item ACC~\cite{yan2021acc}: State-of-the-art solution for ECN tuning based on MARL (not GNN). We have implemented it based on the description in~\cite{yan2021acc}. This solution is designed for online training and it does not perform parameter sharing across agent implementations (i.e., agents are trained independently). For fairness, we re-train the solution on each evaluation scenario selecting the best set of agents after hyper-parameter tuning. 
\end{itemize}

Our goal is to evaluate our trained model in varying DCN scenarios ---most of them different than those seen in training--- and compare the results against these benchmarks. As stated in Section \ref{sec:motivation}, the Flow Completion Time (FCT) is broadly accepted the most complete performance metric in DCNs, and we pay special attention to the 95 and 99 percentiles of this metric due to the aforementioned impact on the partition/aggregate design patterns dominant in today's applications~\cite{alizadeh2010data, li2019hpcc}. Hence, for each scenario, we compute the FCT median, 95-pct and 99-pct obtained by our solution against those of the benchmarks. In particular, we designed a detailed visualization where \textit{i} results are aggregated and shown by flow size; \textit{ii} the FCT values of the baselines are normalized with respect to those of our GraphCC model; and \textit{iii} the flow size distribution is shown in parallel to properly contextualize the relevance of the obtained results. Finally, these plots are complemented with a direct table comparison of the mean absolute values of FCT slowdown, throughput and queue length metrics.

\subsection{Direct Performance Comparison}
\label{subsec:performance}

In our first set of experiments, we aim to evaluate our solution in the same scenario considered in training, with traffic generated from the \emph{FB\_{Hadoop}} workload. We generate two different traffic traces to test the performance of our method: with periodic incasts --similar to those seen during training-- and without them. Respectively, Figures \ref{subfig:fb_incasts} and \ref{subfig:fb_no_incasts} compare the per-flow size FCT median, percentile 95 (p95) and percentile 99 (p99) achieved by GraphCC against the defined baselines. 

In both cases, we can see that GraphCC clearly outperforms the static DCQCN setting and improves the state-of-the-art ACC solution, especially on short and medium sized flows --which represent the vast majority of flows-- in median and both p95 and p99 tails. An aspect that we can observe in this case is that GraphCC learns to slightly sacrifice latency for long flows, as they account for a small percentage of the total counting. Overall, and as shown in Table \ref{tab:metrics_incasts}, GraphCC achieves a notable reduction in the mean FCT slowdown (up to $16.6\%$ reduction w.r.t. ACC without incasts, and $12.6\%$ reduction without) while keeping equivalent mean throughput to that of ACC ($1.38\%$ of difference in the worst case).

Moreover, Table~\ref{tab:metrics_incasts} also shows an interesting behaviour of our solution: it achieves a significant reduction in queue length  --more than $50\%$ in both \emph{FB\_{Hadoop}} experiments--, which directly relates to significantly lower buffer occupancies in switches. These results suggest that GraphCC is able to attain good flow-level latency (i.e., FCT) by learning how to efficiently manage queues so as to achieve stability and be prepared for potential microbursts and incast events. This near-zero queue behavior is a trend seen in other state-of-the-art solutions, such as HPCC~\cite{li2019hpcc} (based on advanced telemetry), and it especially helps achieve ultra-low latency on short flows. As previously discussed in Section~\ref{sec:motivation}, this behavior is also very beneficial in modern DCNs, given the ongoing trend on decreasing the ratio between buffer size and switch capacity.

\subsection{Evaluation under Traffic Changes}
\label{subsec:workloads}

Our next goal is to demonstrate that our trained GraphCC model can adapt to unseen traffic scenarios without requiring any further re-training. In this regard, we design two different challenging scenarios: different workload distributions, and very high traffic loads.

\subsubsection{Different Traffic Workloads}
\label{subsec:workloads}
Different applications may have completely different traffic distributions, so we are interested in testing how our trained GraphCC model performs over workloads not previously seen in training. Hence, in this set of experiments we generate traffic traces from two different real-world traffic distributions, \emph{WebSearch} and \emph{AliStorage}, with and without periodic incasts. Figure \ref{fig:websearch_and_ali} and Table \ref{tab:metrics_incasts} summarize the experimental results. We first notice that the flow size distributions of these workloads greatly differ from the one seen in training --i.e. \emph{FB\_Hadoop}. In particular, \emph{WebSearch}-based traffic (see Figures \ref{subfig:websearch_incasts} and \ref{subfig:websearch_no_incasts}) involves dealing with a considerable higher amount of medium and long-sized flows. However, we can observe that GraphCC improves its FCT metrics for the longer flows on this scenario, suggesting that it is properly prioritizing them. In fact, as shown in Table \ref{tab:metrics_incasts}, GraphCC still gets a slightly better mean FCT slowdown than ACC in this case, with and without incast bursts. On the other hand, GraphCC model again sacrifices a bit of long flows' latency for AliStorage traffic traces (Figures \ref{subfig:ali_incasts} and \ref{subfig:ali_no_incasts}), since in this scenario those flows have even less density than in \emph{FB\_Hadoop} traces. By doing so, our model is able to obtain better FCT median, p95 and p99 metrics than baselines for the major number of flows --with and without incasts--, and reducing the mean FCT slowdown of ACC by more than $11\%$ (Table \ref{tab:metrics_incasts}). As we can also see in Table \ref{tab:metrics_incasts}, GraphCC achieves such good FCT-based results in all these scenarios only at the expense of a small reduction in the mean throughput ($1\%$ of difference w.r.t. ACC at worst). However, GraphCC provides as well with significant improvements in terms of buffer occupancies, lowering ACC's mean queue length from $40-50\%$ for AliStorage traffic, up to more than $80\%$ for \emph{WebSearch} traces.

\subsubsection{Higher Traffic Loads}
\label{subsec:traffic-loads}

\begin{figure}[!t]
    \begin{subfigure}[]{0.49\columnwidth}
	\includegraphics[width=1.0\linewidth]{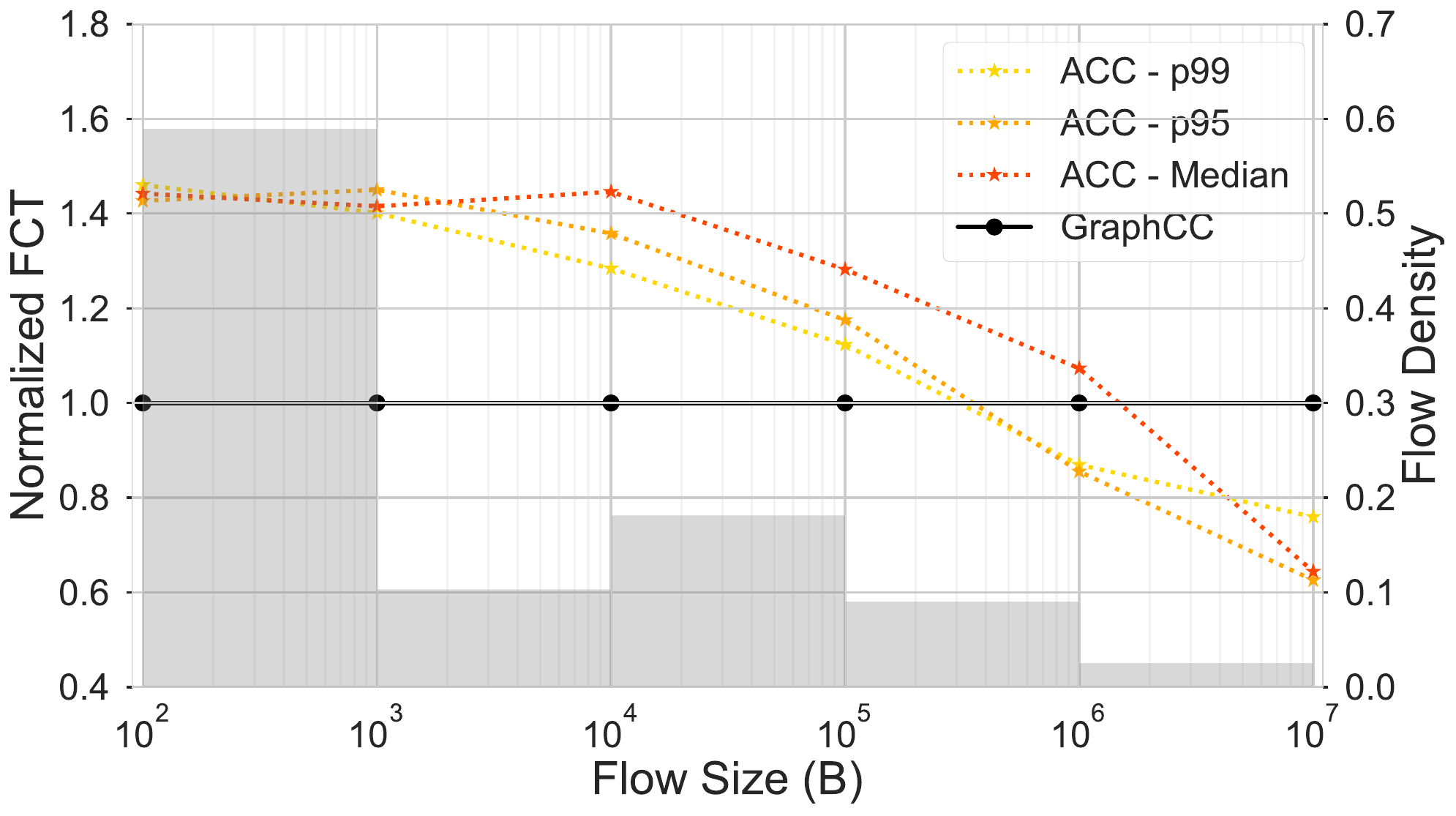}
	\vspace{-0.5cm}
        \caption{Very high load - 70\% load}
	\label{subfig:fb_70_load}
    \end{subfigure}
    \begin{subfigure}[]{0.49\columnwidth}
	\includegraphics[width=1.0\linewidth]{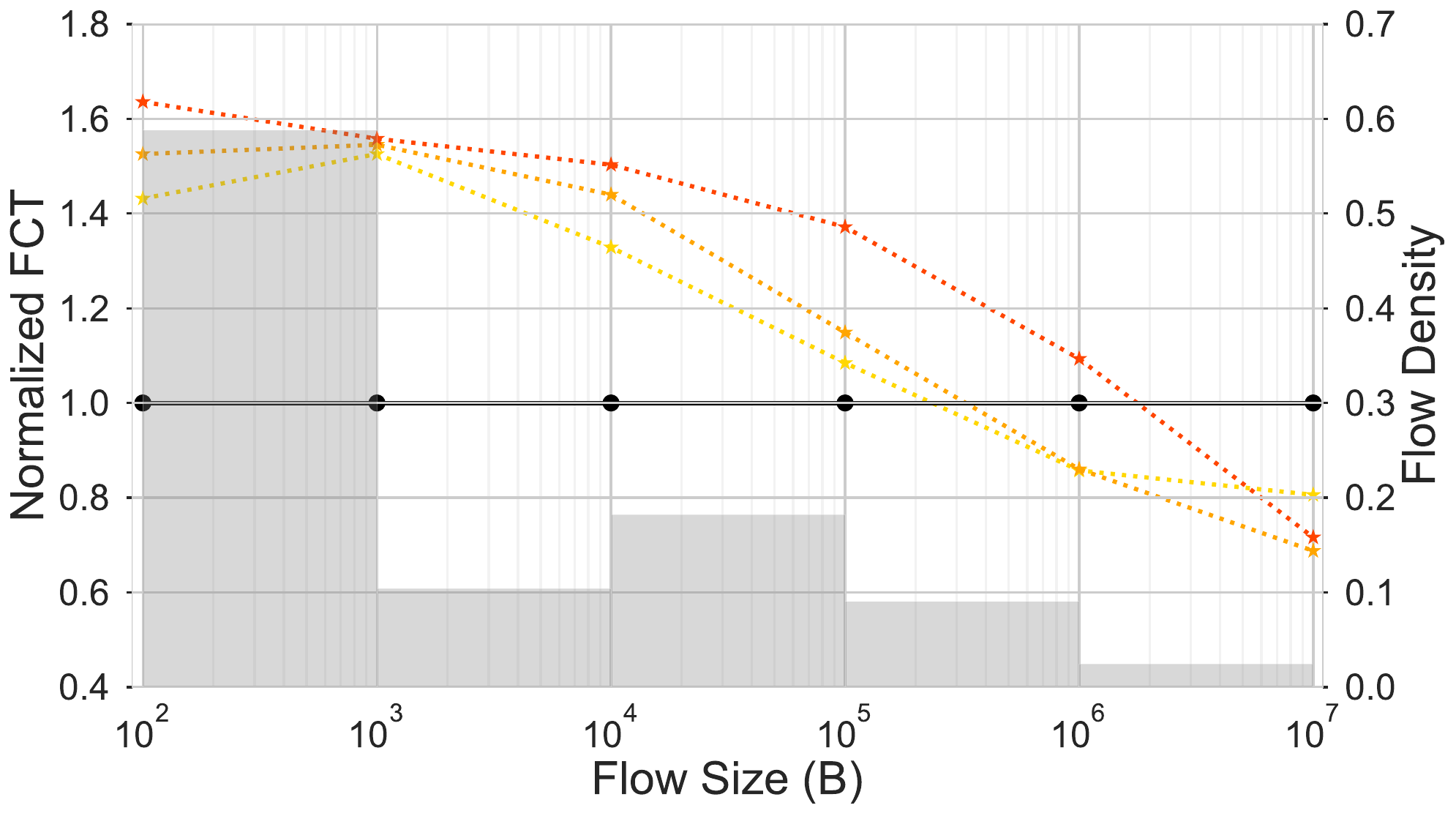}
	\vspace{-0.5cm}
        \caption{Extreme load - 80\% load}
	\label{subfig:fb_80_load}
    \end{subfigure}
     \caption{Evaluation of GraphCC against ACC when considering higher loads than those seen in training (\emph{FB\_Hadoop} workload).}
  \label{fig:fb_loads}
\end{figure}

Next, we analyze whether our trained model can successfully operate over higher traffic loads. For doing so, we evaluate our GraphCC model over new traces from the \emph{FB\_Hadoop} workload with an average network load of 70\% and 80\% --instead of the average 60\% experienced in training. Figure \ref{fig:fb_loads} presents the normalized FCT median, p95 and p99 achieved by ACC with respect to our solution in both cases. We can see both in Figure \ref{fig:fb_loads} and in Table \ref{tab:changes} that the results of both loads are alike, and not very different to those previously observed with a 60\% load: compared to ACC, GraphCC improves all three FCT metrics for short and medium flows --which represent almost 90\% of the total flows--, and achieves equivalent mean throughput --within $1.35\%$. GraphCC manages to reduce the mean queue length by more than 57\%. These results suggest that GraphCC is robust against varying traffic loads.

\begin{table}[!t]
 \resizebox{\columnwidth}{!}{
 \centering
\begin{tabular}{p{1.2cm} c >{\centering\arraybackslash}p{1.1cm} c >{\centering\arraybackslash}p{1.1cm} c >{\centering\arraybackslash}p{1.1cm}}
\toprule
                            &  \multicolumn{2}{c}{\textbf{FCT Slowdown}}  &  \multicolumn{2}{c}{\textbf{Throughput (Mbps)}} &  \multicolumn{2}{c}{\textbf{Queue Length (kB)}}   \\ \cmidrule(lr){2-3} \cmidrule(l){4-5} \cmidrule(l){6-7}
                             &   \textbf{ACC}   & \textbf{GraphCC}  &\textbf{ACC}   & \textbf{GraphCC}  & \textbf{ACC}   & \textbf{GraphCC}    \\ \midrule
1 failure      &  2.91   &   \textbf{2.66} \mbox{(-8.59\%)}  &  \textbf{343}   &  338 \mbox{(-1.46\%)}   &  9.10   &  \textbf{3.89} \mbox{(-57.3\%)}  \\  
2 \mbox{failures}      &  5.80   &  \textbf{5.72} \mbox{(-1.38\%)}   &  \textbf{275}   &  275 \qquad \linebreak \mbox{(=)}    &  7.92   &  \textbf{4.91} \mbox{(-38.0\%)}  \\  \midrule
Extra branch      &  1.76   &  \textbf{1.56} \mbox{(-11.3\%)}   &  \textbf{543}   &  541 \mbox{(-0.37\%)}   &  5.73   & \textbf{2.94} \mbox{(-48.7\%)} \\  \midrule
32 hosts     &  2.74   &  \textbf{2.20} \mbox{(-19.7\%)}   &  \textbf{475}   &  468 \mbox{(-1.47\%)}   &   10.7  &  \textbf{3.80} \mbox{(-64.5\%)} \\
40 hosts    &  2.70   &  \textbf{2.27} \mbox{(-15.9\%)}  &   \textbf{564}  &   560 \mbox{(-0.71\%)} &  8.60   & \textbf{3.65} \mbox{(-57.6\%)} \\
 \bottomrule
\end{tabular}
}
\caption{Mean FCT slowdown, throughput, and queue length of GraphCC and ACC under various topological changes: link failures (top), adding an extra branch (middle), and adding extra hosts (bottom). In all cases, traffic is generated from the \emph{FB\_Hadoop} workload.}
\label{tab:changes}
\end{table}

\subsection{Evaluation under Topology Changes}
\label{subsec:topology-changes}

Lastly, we define a set of experiments involving topological changes to further evaluate the  robustness of a trained GraphCC model. In particular, we analyze two different scenarios: random link failures, and flatter networks --i.e. adding extra hosts connected to leaf switches.

\subsubsection{Link failures}

\begin{figure}[!t]
    \begin{subfigure}[]{0.49\columnwidth}
	\includegraphics[width=1.0\linewidth]{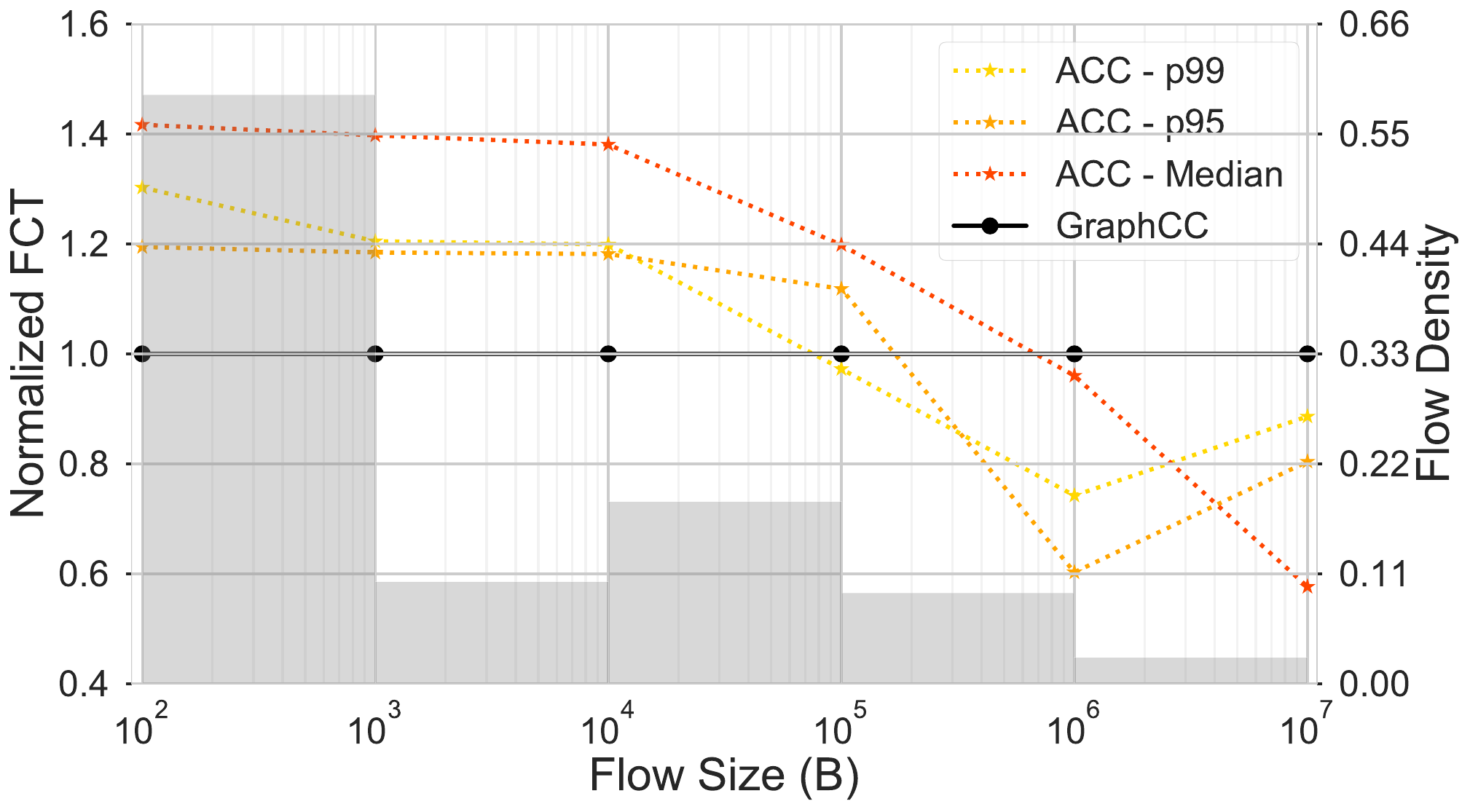}
	\vspace{-0.5cm}
        \caption{1 link failure}
	\label{subfig:fb_1_link}
    \end{subfigure}
    \begin{subfigure}[]{0.49\columnwidth}
	\includegraphics[width=1.0\linewidth]{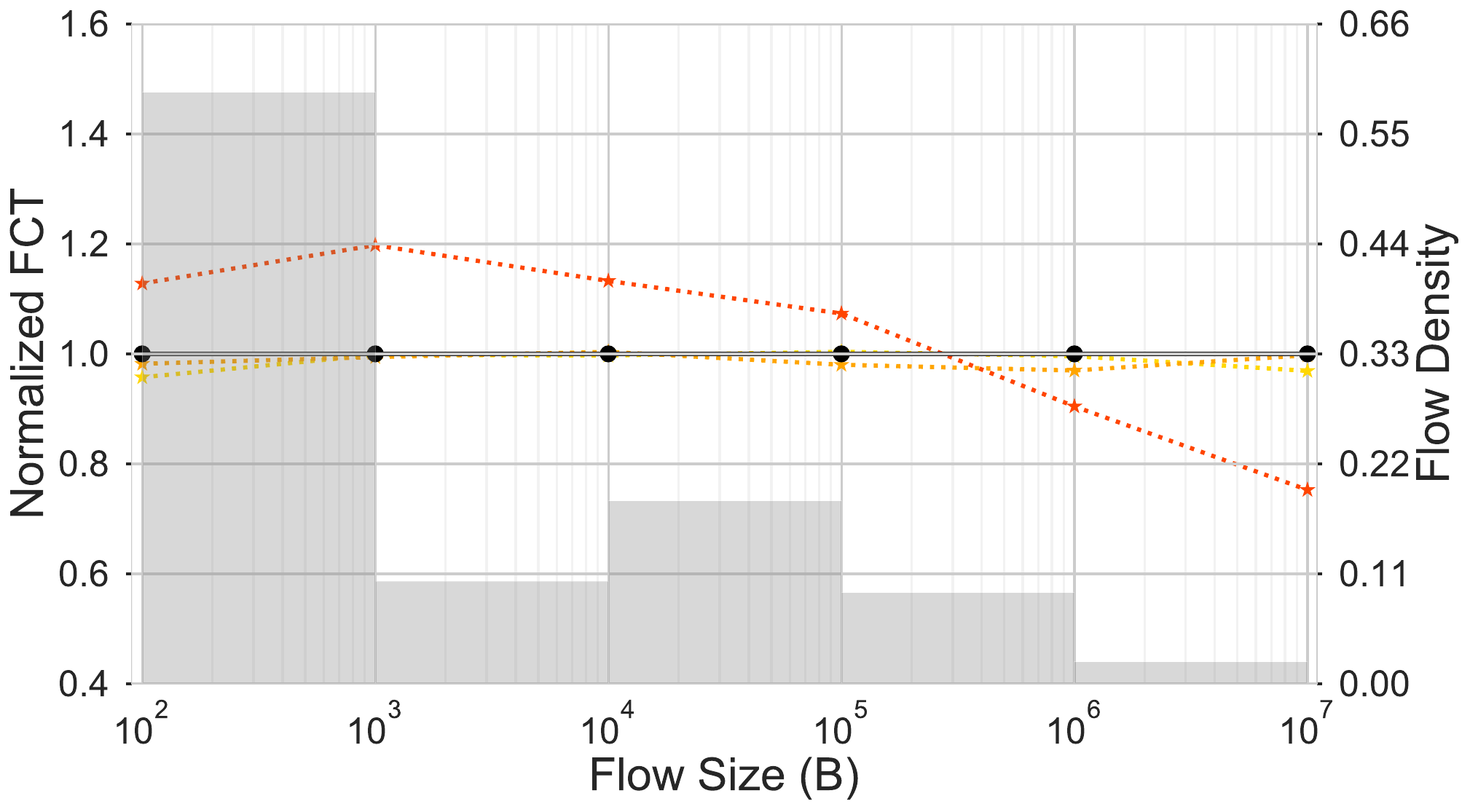}
	\vspace{-0.5cm}
        \caption{2 link failures}
	\label{subfig:fb_2_links}
    \end{subfigure}
     \caption{Evaluation of GraphCC against ACC when the DCN suffers from link failures (\emph{FB\_Hadoop} workload).}
  \label{fig:fb_links}
\end{figure}

In production DCNs, link failures occur frequently~\cite{zhou2014wcmp, liu2013f10}, and often lead to network asymmetries and severe performance degradation (see Section \ref{subsec:dynamic-cc}). In these experiments, we simulate how GraphCC responds to critical failures in links between leaf and spine switches. Figures~\ref{subfig:fb_1_link} and \ref{subfig:fb_2_links} show, respectively, a comparison of FCT metrics between GraphCC and ACC, for 1 and 2 link failures. In the case of a single failure, GraphCC improves ACC on these metrics for most of the flows. On the other hand, the 2-link failure scenario presents less performance gap between our method and ACC, which was individually trained on each target network scenario. Here, GraphCC gets a small improvement in median FCT for small and medium flows, but the p95 and p99 tails are equivalent to those obtained by ACC for all sizes. This exposes how challenging this scenario is, especially for models pre-trained offline, as it is the case of GraphCC. Overall, in terms of absolute values (Table \ref{tab:changes}), even with 2 failures the proposed solution still provides with a slightly better mean FCT slowdown serving exactly the same throughput, and does so while reducing the mean queue length up to a 38\%.

\subsubsection{Network upgrade} 
Datacenters are periodically upgraded to increase their processing and switch capacity, for example by adding new Points-of-Delivery to the network~\cite{yan2021acc}. This can change drastically the overall network state. In these experiments we aim to simulate a topology upgrade. In particular, we add a new branch to the 2-layer Clos topology considered in our experiments, which includes an additional core switch, two spine switches, and six hosts. 
In Table~\ref{tab:changes} we observe that GraphCC achieves significant improvements in terms of FCT w.r.t. ACC (11.3\%), while keeping the same throughput and considerable queue length reduction (48.7\%).

\subsubsection{Adding extra hosts} 
\begin{figure}[!t]
    \begin{subfigure}[]{0.49\columnwidth}
	\includegraphics[width=1.0\linewidth]{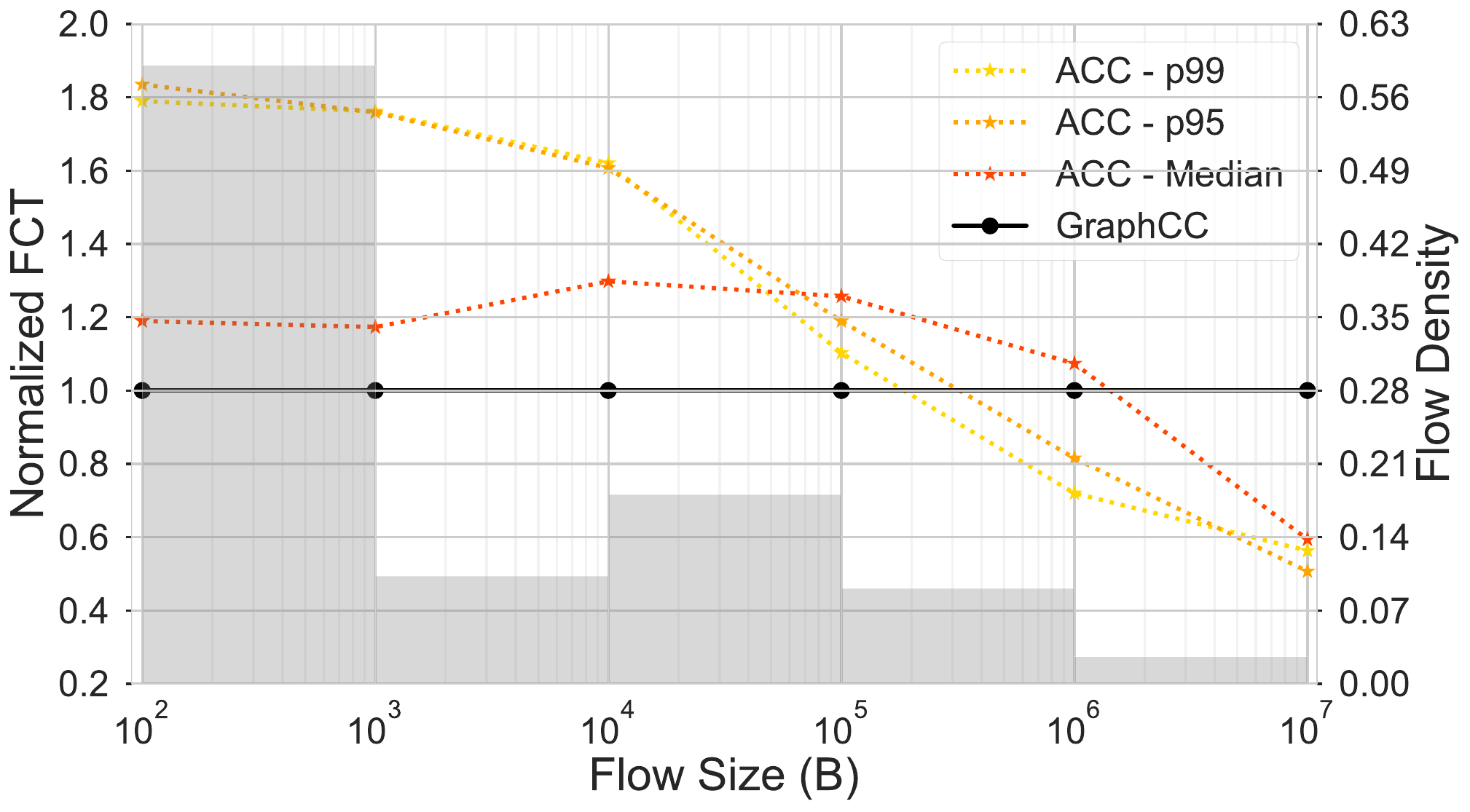}
	\vspace{-0.5cm}
        \caption{32 hosts (8:1 ratio)}
	\label{subfig:fb_16_hosts}
    \end{subfigure}
    \begin{subfigure}[]{0.49\columnwidth}
	\includegraphics[width=1.0\linewidth]{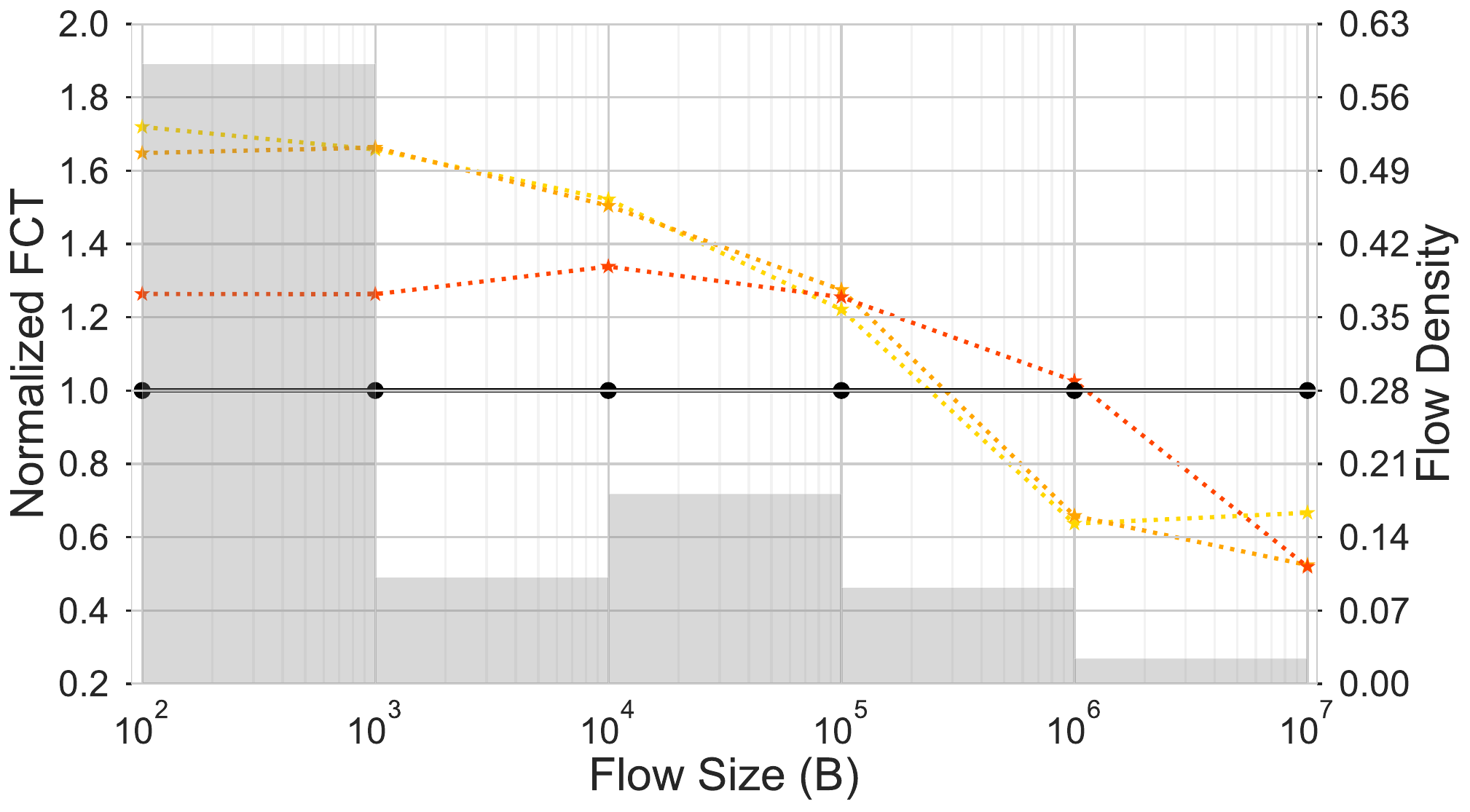}
	\vspace{-0.5cm}
        \caption{40 hosts (10:1 ratio)}
	\label{subfig:fb_20_hosts}
    \end{subfigure}
     \caption{Evaluation of GraphCC against ACC on flatter networks (\emph{FB\_Hadoop} workload).}
  \label{fig:fb_hosts}
\end{figure}

Datacenters are periodically upgraded to increase their processing and switch capacity, for example by adding new Points-of-Delivery to the network~\cite{yan2021acc}. This can change drastically the overall network state. In this section, we aim to evaluate how GraphCC operates when it is deployed in network topologies with different properties. 
For this purpose, we increasingly connect new hosts to leaf switches in the original network where GraphCC was trained (with 24 hosts, i.e. 6:1 host/switch ratio). Figure~\ref{fig:fb_hosts} gathers a comparison of the FCT between GraphCC and ACC for two new scenarios (with 32 and 40 hosts). Likewise, Table~\ref{tab:changes} shows the corresponding aggregated mean values. The overall performance comparison w.r.t ACC --FCT, throughput and queue length metrics-- is similar to that obtained for the original topology with 24 hosts. We observe an increment on the mean throughput --due to the increase of hosts--, as well as a very slight behaviour difference on the p95 and p99 tails of the FCTs for small flows, if we compare it with the analogous results in the original network (Figure~\ref{subfig:fb_no_incasts}). Overall, we see that GraphCC improves ACC a bit further on these metrics, which suggests that our solution can effectively handle flatter networks with higher congestion levels on the core.

\section{Related Work}\label{sec:related-work}
CC has been largely studied in the past, with a rich body of proposals especially focused on high-speed DCNs. This section comprises an overview of relevant works related to GraphCC.

\vspace{0.1cm}
\noindent \textbf{Advanced CC mechanisms for DCNs:} Some recent pioneering works have proposed sophisticated CC mechanisms showing outstanding performance in DCNs~\cite{alizadeh2010data,zhu2015congestion,mittal2015timely,montazeri2018homa,li2019hpcc,kumar2020swift}. For example, HPCC~\cite{li2019hpcc} achieves remarkably short FCTs while offering good throughput and stability to traffic changes (e.g., incast events). To this end, it leverages accurate fine-grained measurements produced by modern In-band Network Telemetry (INT) mechanisms~\cite{tan2021band, ben2020pint}. TIMELY~\cite{mittal2015timely} and Swift~\cite{kumar2020swift} rely on accurate delay measurements on NICs to control flow rates. Other well-known works, such as pHost~\cite{gao2015phost} or Homa~\cite{montazeri2018homa} are credit-based solutions, where receivers control the flow rate by sending credit packets to senders. All the previous CC mechanisms rely on novel network architectures and/or protocol stacks that unfortunately are not supported by most legacy switches deployed in DCNs nowadays. At the time of this writing, the most widely deployed CC standards are DCTCP~\cite{alizadeh2010data} in networks running the TCP/IP stack, and DCQCN~\cite{zhu2015congestion} in RDMA-based networks. Both are ECN-based mechanisms, where switches mark packets in case they detect congestion (i.e., queue length above certain thresholds). GraphCC is designed to inter-operate with traditional ECN-based mechanisms in an efficient and distributed way. Although in this paper we test GraphCC only in RDMA networks with DCQCN~\cite{zhu2015congestion}, our solution can be easily adapted to optimize any other ECN-based CC mechanism, such as the aforementioned DCTCP~\cite{alizadeh2010data}, or TCP-Bolt~\cite{stephens2014practical}).

\noindent \textbf{Machine Learning-based mechanisms for CC:} Recent works posit the use of ML techniques to produce data-driven solutions that can efficiently adapt to the network dynamics. Among the most popular solutions, works such as Aurora~\cite{jay2019deep}, or Orca~\cite{abbasloo2020classic}, propose to use RL to adapt flow rates at end-hosts. These solutions are focused on adapting the flow rate at end-hosts according to the congestion feedback received, and they require to re-implement the network stack. Likewise, AuTO~\cite{chen2018auto} proposes a novel two-level mechanism that accurately controls routing for long flows and queue scheduling for optimizing the latency of short flows. Instead, GraphCC is focused on distributed in-network optimization of widely deployed ECN-based CC mechanisms. The closest work to GraphCC is arguably ACC~\cite{yan2021acc}, where the authors propose a MARL-based mechanism to optimize ECN-based schemes. However, that work ---~and all previous ones~--- contemplate online training to dynamically learn how to adapt to the network state. As a result, these ML-based solutions may suffer from transient performance degradations when changes occur in the network (e.g., new traffic workload, incast events, failures). Also, online training entails an intrinsic uncertainty on what would be the resulting performance after re-training. This would require strong supervision mechanisms to check the evolution of agents and be ready to deploy backup mechanisms. 

On the other hand, GraphCC naturally exhibits high robustness to generalize across traffic and topology changes unseen during the training phase. This is thanks to the GNN architecture that it internally implements, which naturally induces a cooperation mechanism among agents ---via message exchanges between neighbours-- that contrasts with the greedy behaviour of ACC agents ---which act based on its local measurements, as no contextual data is provided to them. On top of that, our GNN design implements \emph{parameter sharing} to finally produce a general agent implementation jointly learned from the individual perspective of each node in the network. This property can be especially interesting from a practical standpoint, as it enables to train the solution offline (e.g., in a controlled testbed), and then be able to deploy it directly in production networks, without the need for (re)training it on premises. Also, it permits extensive testing prior to deployment, giving vendors the possibility to issue certifications with the safe operational ranges that the solution would safely support once deployed (e.g., link capacities, max. network size). This process would be better aligned with the standard way network products are commercialized nowadays.

\vspace{-0.1cm}

\section{Conclusion}\label{sec:conclusion}

This paper has introduced GraphCC, a distributed solution for in-network CC optimization in DCNs. GraphCC is compatible with any network running widely deployed ECN-based CC mechanisms, such as DCQCN, or DCTCP. In GraphCC, agents are deployed in switches; they cooperate and exchange information to dynamically adapt the ECN configuration and optimize the global flow-level performance. In our evaluation, we have benchmarked GraphCC against two baselines: $(i)$ a static ECN configuration used in Alibaba's production DCNs, and $(ii)$ ACC, a state-of-the-art ML-based solution for dynamic ECN tuning. The experimental results show that our solution significantly outperforms the two previous baselines in terms of Flow Completion Time. At the same time, we have observed that our solution learns to keep small queue lengths (up to 85.7\% w.r.t. ACC), thus being especially interesting for next-generation DCNs, where buffer size is expected to continue shrinking w.r.t. the skyrocketing switch capacities.

As discussed earlier, an important feature of ML-based solutions for networking is their capability to generalize to different scenarios to those seen during training, as it avoids the need for online training. However, existing ML-based CC optimization solutions are designed to be trained online and gradually learn how to adapt to the current network conditions. In this vein, GraphCC exhibits good behavior when operating in new scenarios never seen during training, such as shifts on the traffic workload, or topology changes. This is thanks to its internal GNN-based framework, which leverages two main features: $(i)$ during training GraphCC produces a single agent implementation jointly learned from the individual perspective of each agent in the network, using \emph{parameter sharing}, and $(ii)$ agents run a topology-aware message passing mechanism to get local context and cooperate with each other. As a result, GraphCC produces more robust and general agent implementations that can successfully operate on significantly different network scenarios to those seen during training.  

\vspace{0.25cm}

\noindent \textbf{Acknowledgments:} This publication is part of the Spanish I+D+i project TRAINER-A (ref.~PID2020-118011GB-C21), funded by MCIN/ AEI/10.13039 /501100011033. This work is also partially funded by the Catalan Institution for Research and Advanced Studies (ICREA), the Secretariat for Universities and Research of the Ministry of Business and Knowledge of the Government of Catalonia, and the European Social Fund.

\bibliographystyle{ACM-Reference-Format}
\bibliography{reference}

\end{document}